\pdfoutput=1

\documentclass[11pt,twoside,a4paper,cmspaper,final,collab]{cms-tdr}
\def\svnVersion{258627}\def\cmsCernNo{2013-037}\def\cmsCernDate{\today}\def\cmsMessage{Published in the Journal of High Energy Physics as \href{http://dx.doi.org/10.1007/JHEP08(2014)173}{\doi{10.1007/JHEP08(2014)173.}}}

\begin{document}\cmsNoteHeader{EXO-12-024}

\hyphenation{had-ron-i-za-tion}
\hyphenation{cal-or-i-me-ter}
\hyphenation{de-vices}

\RCS$Revision: 256732 $
\RCS$HeadURL: svn+ssh://svn.cern.ch/reps/tdr2/papers/EXO-12-024/trunk/EXO-12-024.tex $
\RCS$Id: EXO-12-024.tex 256732 2014-08-19 15:19:14Z alverson $
\newlength\cmsFigWidth
\ifthenelse{\boolean{cms@external}}{\setlength\cmsFigWidth{0.85\columnwidth}}{\setlength\cmsFigWidth{0.6\textwidth}}
\ifthenelse{\boolean{cms@external}}{\providecommand{\cmsLeft}{top}}{\providecommand{\cmsLeft}{left}}
\ifthenelse{\boolean{cms@external}}{\providecommand{\cmsRight}{bottom}}{\providecommand{\cmsRight}{right}}
\providecommand{\re}{\ensuremath{\cmsSymbolFace{e}}}
\renewcommand{\GeVc}{\GeV}
\renewcommand{\TeVc}{\TeV}
\renewcommand{\GeVcc}{\GeV}
\renewcommand{\TeVcc}{\TeV}
\newcommand{\intlumi}{19.7\fbinv}
\newcommand{\scalefactorHP}{\ensuremath{0.86 \pm 0.07}}
\newcommand{\scalefactorLP}{\ensuremath{1.39 \pm 0.75}}
\newcommand{\scalefactorHPuNoPer}{\ensuremath{7.5}}
\newcommand{\scalefactorLPuNoPer}{\ensuremath{54}}
\newcommand{\scalefactorHPu}{\ensuremath{7.5\%}}
\newcommand{\scalefactorLPu}{\ensuremath{54\%}}
\newcommand{\GRS}{\ensuremath{\mathrm{G}_\mathrm{RS}}\xspace}
\newcommand{\GBulk}{\ensuremath{\mathrm{G}_\mathrm{bulk}}\xspace}
\newcommand{\MPl}{\ensuremath{\overline{M}_\text{Pl}}\xspace}
\newcommand{\ptk}{\ensuremath{p_{\mathrm{T},k}}\xspace}
\cmsNoteHeader{EXO-12-024} % This is over-written in the CMS environment: useful as preprint no. for export versions
\title{Search for massive resonances in dijet systems containing jets tagged as W or Z boson decays in pp collisions at $\sqrt{s}= 8\TeV$}

\date{\today}

\abstract{A search is reported for massive resonances decaying into a
  quark and a vector boson (W or Z), or two vector bosons
  (WW, WZ, or ZZ).  The analysis is
  performed on an inclusive sample of multijet events corresponding to
  an integrated luminosity of 19.7\fbinv, collected in
  proton-proton collisions at a centre-of-mass energy of 8\TeV with
  the CMS detector at the LHC. The search uses novel jet-substructure
  identification techniques that provide sensitivity to the presence of
  highly boosted vector bosons decaying into a pair of
  quarks. Exclusion limits are set at a confidence level of 95\% on
  the production of: (i) excited quark resonances $\Pq^*$ decaying to
  qW and qZ for masses less than 3.2\TeV and
  2.9\TeV, respectively, (ii) a Randall--Sundrum graviton \GRS decaying
  into WW for masses below 1.2\TeV, and (iii) a heavy
  partner of the W boson $\PWpr$ decaying into WZ for
  masses less than 1.7\TeV.
For the first time mass limits are set on $\PWpr\to \PW\cPZ$ and
$\GRS\to \PW\PW$
in the all-jets final state. The mass limits on
$\Pq^*\to \Pq\PW$, $\Pq^*\to \Pq\cPZ$, $\PWpr\to \PW\cPZ$,
$\GRS\to \PW\PW$ are the most stringent to date.
A model with a ``bulk'' graviton \GBulk that decays
into WW or ZZ bosons is also studied.}

\hypersetup{%
 pdfauthor={CMS Collaboration},%
 pdftitle={Search for massive resonances in dijet systems containing jets tagged as W or Z boson decays in pp collisions at sqrt(s)= 8 TeV},%
 pdfsubject={CMS},%
 pdfkeywords={CMS, physics, dijet, jet substructure, resonances}}

\maketitle

\section{Introduction}
\label{sec:introduction}

Several models of physics beyond the standard model (SM) predict the
existence of resonances with masses above 1\TeV that decay into a
quark and a \PW\ or \cPZ\ vector boson, or into two vector bosons. In
proton-proton (pp) collisions at the energies reached at the Large Hadron
Collider (LHC), vector bosons emerging from such decays usually would have
sufficiently large momenta so that the hadronization products of their
\cPaq \Pq(') decays would merge into a single massive
jet~\cite{Gouzevitch:2013qca}. We present a search for events
containing one or two jets of this kind in pp collisions at a
centre-of-mass energy of $\sqrt{s}=8$\TeV.  The data sample,
corresponding to an integrated luminosity of 19.7\fbinv, was collected
with the CMS detector at the LHC.

The signal is characterized by a peak in the dijet invariant mass
distribution $m_\mathrm{jj}$ over a continuous background from SM
processes, comprised mainly of multijet events from quantum
chromodynamic (QCD) processes. The sensitivity to jets from \PW\ or
\cPZ\ bosons is enhanced through the use of jet-substructure
techniques that help differentiate such jets from remnants of quarks
and gluons~\cite{topwtag_pas,JME-13-006}, providing the possibility of
``$\PW/\cPZ$-tagging''. This search is an update of a previous CMS
study~\cite{ref_2011} performed using data from pp collisions at
$\sqrt{s}=7$\TeV. Besides increased data-sample size and larger
signal cross sections from the increase in centre-of-mass energy, this
analysis also benefits from an improved $\PW/\cPZ$-tagger based on
``$N$-subjettiness'' variables, introduced in
Ref.~\cite{Thaler:2010tr} and defined in Section~\ref{sec:analysis}.

We consider four reference processes that yield one $\PW/\cPZ$-tagged
or two $\PW/\cPZ$-tagged all-jet events: (i) an excited quark
$\Pq^*$~\cite{ref_qstar,ref_qstar2} that decays into a quark and
either a \PW\ or a \cPZ\ boson, (ii) a Randall--Sundrum (RS) graviton
$\GRS$ that decays into \PW \PW\ or \cPZ \cPZ\
bosons~\cite{rs1,Randall:1999vf}, (iii) a ``bulk'' graviton $\GBulk$
that decays into \PW \PW\ or \cPZ \cPZ\
~\cite{GravitonWWZZ1,GravitonWWZZ2,GravitonWWZZ3}, and (iv) a
heavy partner of the SM \PW\ boson (\PWpr) that decays into \PW\cPZ\
\cite{egm}.

Results from previous searches for these signal models include limits
placed on the production of $\Pq^*$ at the LHC as
dijet~\cite{exo12016,ATLASexcitedPAS,Harris:2011bh} or
$\gamma$+jet~\cite{Aad:2013cva} resonances, with a $\Pq^*$ lighter
than $\approx$3.5\TeVcc at a confidence level (CL) of
95\%~\cite{exo12016}. Specific searches for resonant $\Pq\PW$ and
$\Pq\cPZ$ final states at the
Tevatron~\cite{CDFexcitedPAPER,D0excitedPAPER} exclude $\Pq^*$ decays
into $\Pq\PW$ or $\Pq\cPZ$ with $m_{\Pq^*}<0.54\TeVcc$, and results
from the LHC~\cite{ref_2011,CMSqZPAS} exclude $\Pq^*$ decays into
$\Pq\PW$ or $\Pq\cPZ$ for $m_{\Pq^*}< 2.4\TeVcc$ and $m_{\Pq^*}<
2.2\TeVcc$, respectively.

Resonances in final states containing candidates for \PW\PW\ or
\cPZ\cPZ\ systems have also been
sought~\cite{CMSZZPAS2,ATLASWWPAPER,ATLASZZPAPER,CDFZZPAPER}, with
lower limits set on the masses of $\GRS$ and $\GBulk$ as a function of
the coupling parameter $k/\MPl$, where $k$ reflects the curvature of
the warped space, and $\MPl$ is the reduced Planck mass ($\MPl \equiv
M_\text{Pl}/\sqrt{8\pi}$)~\cite{rs1,Randall:1999vf}. The bulk graviton
model is an extension of the original RS model that addresses the
flavour structure of the SM through localization of fermions in the
warped extra dimension. The experimental signatures of the $\GRS$ and
$\GBulk$ models differ in that $\GBulk$ favours the production of
gravitons through gluon fusion, with a subsequent decay into vector bosons,
rather than production and decay through fermions or photons, as the
coupling to these is highly suppressed. As a consequence, $\GBulk$
preferentially produces \PW\ and \cPZ\ bosons that are longitudinally polarized,
while $\GRS$ favours the production of transversely polarized
\PW\ or \cPZ\ bosons. In this study, we use an improved calculation of
the $\GBulk$ production cross section~\cite{GravitonWWZZ1,
  CarvalhoThesis2014} that predicts a factor of four smaller yield than
assumed in previous studies~\cite{CMSZZPAS2,ATLASWWPAPER}.

The most stringent limits on \PWpr boson production  are those reported
for searches in leptonic final
states~\cite{CMSwprimePAPER2013,ATLASwprimePAPER}, with the current
limit specified by $m_{\PWpr}>2.9\TeVcc$. Depending on the chirality
of the $\PWpr$ couplings, this limit could change by $\approx$0.1\TeVcc. Searches for $\PWpr$ in the \PW\cPZ\ channel have also been
reported~\cite{CMSwprimeWZPAS,ATLASWWPAPER,ATLASwprimeWZPAS} and set
a lower limit of $m_{\PWpr}>1.1\TeVcc$.

The CMS detector, the data, and the event simulations are described briefly
in Section~\ref{sec:cms_detector}. Event reconstruction, including
details of $\PW/\cPZ$-tagging, and selection criteria are discussed in
Section~\ref{sec:analysis}. Section~\ref{sec:background} presents
studies of dijet mass spectra, including SM background estimates.
The systematic uncertainties are discussed in Section~\ref{sec:systematics},
the interpretation of the results in terms of the benchmark signal models
is presented in Section~\ref{sec:results}, and the results are summarized in Section 7.

\section{The CMS detector, data, and simulated event samples}
\label{sec:cms_detector}

The CMS detector~\cite{:2008zzk} is well-suited to reconstructing
particle jets, as it contains highly segmented electromagnetic and
hadronic calorimeters and a fine-grained precision tracking
system. Charged-particle trajectories are reconstructed in the inner
silicon tracker, which consists of a pixel detector surrounded by
silicon strip detectors and is immersed in a 3.8\unit{T} magnetic
field. A lead tungstate crystal electromagnetic calorimeter (ECAL) and
a brass/scintillator hadron calorimeter (HCAL) surround the tracking
volume, and provide complementary information for reconstructing
photons, electrons, and jets. Muon trajectories are measured in
gas ionization detectors embedded in the outer steel return yoke of the
CMS magnet.

CMS uses a coordinate system with the origin located at the nominal
collision point, the $x$ axis pointing towards the centre of the LHC
ring, the $y$ axis pointing up (perpendicular to the plane containing
the LHC ring), and the $z$ axis along the counterclockwise beam
direction. The azimuthal angle, $\phi$, is measured with respect to
the $x$ axis in the $(x,y)$ plane, and the polar angle, $\theta$, is
defined with respect to the $z$ axis. The tracker covers the full
azimuthal range of $0 \le \phi < 2\pi$ within $\abs{\eta} < 2.5$, where
$\eta$ is the pseudorapidity defined as $\eta =
-\ln[\tan(\theta/2)]$. The coverages of the ECAL and HCAL extend to
$\abs{\eta} < 3$ and $\abs{\eta} < 5$, respectively. The calorimeter cells are
grouped into towers projecting radially outward from the centre of the
detector. In the central region ($\abs{\eta}<1.74$) the towers have
dimensions $\Delta\eta = \Delta\phi = 0.087$ radians, and these
increase with $\abs{\eta}$ in the forward regions.

The signals of interest are simulated using \textsc{jhugen}~\cite{Gao:2010qx,Bolognesi:2012mm},
\PYTHIA~6.426~\cite{pythia}, and \HERWIG{++} 2.5.0~\cite{herwig} Monte
Carlo (MC) event generators, and processed through a simulation of the
CMS detector, based on \GEANTfour~\cite{refGEANT}. Tune
Z2*~\cite{bib_tunez1}
is used in \PYTHIA, while the version 23 tune~\cite{herwig} is used in
\HERWIG{++}. The CTEQ61L~\cite{cteq} parton distribution functions
(PDF) are used for \PYTHIA and the MRST2001~\cite{mrst} leading-order
(LO) PDF for \HERWIG{++}. The $\Pq^*\to\PW$+jet and $\cPZ$+jet
processes are generated using \PYTHIA.

The RS graviton production is
studied for $k/\MPl=0.1$, which sets the resonance widths at
$\approx$1\% of the resonance mass, a factor of five smaller than the
experimental resolution in $m_\mathrm{jj}$ for resonance masses
considered in the analysis.
RS graviton cross sections from \PYTHIA are used in the analysis to maintain
consistency in comparisons with related
studies~\cite{CMSZZPAS2}. On the other hand, \HERWIG{++} contains a
more precise description of the angular distributions for $\GRS$
production than \PYTHIA~\cite{resonanceshape}, and is therefore used
to model the $\GRS$ signal.

Bulk graviton events are generated with $k/\MPl$ ranging from 0.1 to
0.5. Due to the detector resolution on the $m_\mathrm{jj}$ peak, the increase
of the resonance width with $k/\MPl$ has no impact on the signal
distribution for $k/\MPl$ values in the considered range. The
reference samples are generated assuming $k/\MPl=0.2$, with
\textsc{jhugen} interfaced to \PYTHIA for the showering and
hadronization of quarks. Bulk graviton production is studied up to
$k/\MPl=0.5$, where the resonance width is $\approx$1\% of the
resonance mass.  The $\PWpr\to\PW\cPZ$ process is generated using
\PYTHIA, assuming SM $\mathrm{V}-\mathrm{A}$ couplings~\cite{pythia}.
The cross section values are scaled to next-to-next-to-leading order
(NNLO) values with the K-factors obtained using the simulation code
\textsc{fewz} 2.0~\cite{wprimekfactor}.

All simulated samples are passed through the standard CMS event
reconstruction software. Data are compared to simulated samples of
multijet events, generated using both \HERWIG{++} and {\MADGRAPH
  5v1.3.30}~\cite{madgraph}, and interfaced to \PYTHIA for parton
showering and hadronization. The simulated sample of multijet events
serves only to provide guidance and cross-checks, as the distribution
of the background is modelled from data.

\section{Event reconstruction and selections}\label{sec:analysis}

In this study the event selection, in the online trigger as well as
offline, utilizes a global view of the event involving information
combined from the individual subdetectors.  Online, events are
selected by at least one of two specific triggers, one based on the
scalar sum of the transverse momenta \pt of the jets, and the other on
the invariant mass $m_\mathrm{jj}$ of the two jets with highest
\pt. The offline reconstruction, described below, is also based on a
global event description.

Events must have at least one reconstructed vertex with $\abs{z} <
24\unit{cm}$. The primary vertex is defined as the one with the
largest summed $\pt^2$ of its associated tracks. Individual particles
are reconstructed and identified using the particle-flow (PF)
algorithm~\cite{particleflow,particleflow2}, and divided into five
categories: muons, electrons, photons (including their conversions
into $\Pep\Pem$ pairs), charged hadrons, and neutral hadrons. Charged
PF candidates not originating from the primary vertex are discarded,
which reduces the contamination from additional pp interactions in the
same or neighbouring bunch crossings (pileup). Ignoring isolated
muons, jets are clustered from the remaining PF candidates using the
Cambridge--Aachen (CA)~\cite{CAaachen,CAcambridge} jet clustering
algorithm, as implemented in
\textsc{FastJet}~\cite{fastjet1,fastjet}. A distance parameter $R=0.8$
is used for the CA algorithm. An event-by-event correction based on
the jet area
method~\cite{jetarea_fastjet,jetarea_fastjet_pu,JME-JINST} is applied
to remove the remaining energy deposited by neutral particles
originating from other interaction vertices. The pileup-subtracted jet
four-momenta are then corrected to account for the difference between
the measured and true energies of hadrons~\cite{JME-JINST}. Finally,
events with jets originating from calorimeter noise are rejected,
requiring that a fraction of the jet energy is also detected in the
tracking system.  Following this selection, the jet reconstruction
efficiencies (estimated from simulation) are larger than 99.9\%, and
contribute negligibly to the systematic uncertainties for signal
events.

Events are selected by requiring at least two jets with $\pt >
30\GeVc$ and $\abs{\eta} < 2.5$. The two jets of highest $\pt$ are
required to have a pseudorapidity separation $\abs{\Delta\eta}<1.3$ to
reduce background from multijet events~\cite{cmsdijet}. The
invariant mass of the two selected jets is required to have
$m_\mathrm{jj}> 890\GeVcc$, which leads to a 99\% trigger efficiency,
with a negligible systematic uncertainty.

\begin{figure}[thb]
\begin{center}
\includegraphics[width=0.49\textwidth]{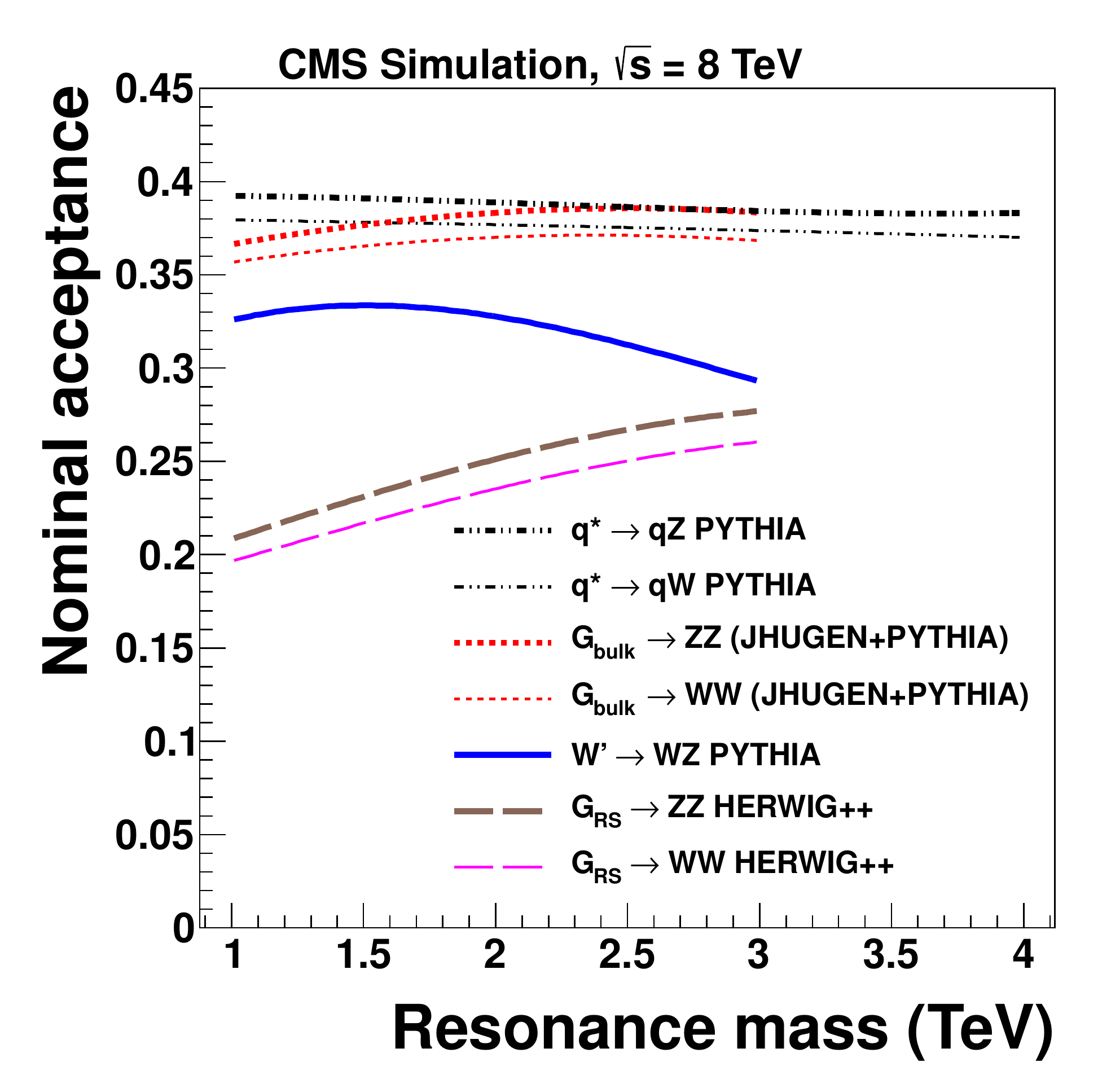}
\end{center}
\caption{The fraction of simulated signal events expected for vector
  bosons decaying into two quarks, reconstructed as two jets, that
  pass the geometrical acceptance criteria ($\abs{\eta} < 2.5$,
  $\abs{\Delta\eta}<1.3$), shown as a function of the dijet invariant
  mass.\label{fig:acceptance}}
\end{figure}

The event selection efficiency for signal is estimated using fully simulated
signal event samples, as described below. These studies also show that less than 1\% of the
events decaying to $\PW\PW$ or $\cPZ\cPZ$ that pass the event
selection criteria are from $\PW\PW\to\ell \cPgn_\ell \cPaq \Pq'$ or
$\cPZ\cPZ\to\ell^+ \ell^-\cPaq \Pq$ decays, where $\ell$ refers to a
muon or an electron. Further, less than 1\% of the selected $\PW\PW$
events are from $\PW\PW \to \Pgt\Pgngt \cPaq \Pq'$ decays, and only 3\% of the
selected $\cPZ\cPZ$ events correspond to $\cPZ\cPZ\to\Pgt^+\Pgt^-
\cPaq \Pq$ decays. Hence, these contaminants are negligible and the
event selection efficiency is dominated by the final states where the
$\PW$ and $\cPZ$ bosons decay to quarks.

Although we use a full simulation to derive the exclusion limits, to
enable reinterpretation of the results in models with other
acceptances, in the following we consider the global efficiency
approximated by the product of ``nominal acceptances'' and the
$\PW/\cPZ$ tagging efficiency, restricted to final states where the
$\PW$ or $\cPZ$ boson decay to quarks. A matching is required within
$\Delta R = \sqrt{(\Delta \eta)^2 + (\Delta\phi)^2} <0.5$ of the
generated \PW\ and \cPZ\ bosons decaying to quarks and their
reconstructed single jets, as part of the nominal acceptances.  The
product of nominal acceptances and the $\PW/\cPZ$ tagging efficiency,
ignoring leptonic decays and the correlations between detector
acceptance and $\PW/\cPZ$ tagging, agree to better than $10\%$ with
the full simulations.

In the analysis reported in this paper, the global efficiency is
estimated from the full simulation of signal events, without applying
the matching requirement.  In this way, the correlations between the
acceptance and $\PW/\cPZ$-tagging efficiency are properly taken into
account.  However, interpreting this search in terms of these nominal
acceptances and$\PW/\cPZ$-tagging efficiencies for any arbitrary model
requires the implementation of an additional uncertainty of 10\%.

The nominal acceptance, shown in Fig.~\ref{fig:acceptance} as a
function of the dijet resonance mass for several signals, takes into
account the angular acceptance ($\abs{\eta} < 2.5$, $\abs{\Delta\eta}<1.3$),
the matching, and the branching fraction into quark final states. The
acceptance for the $\GRS$ model is lower than for the $\GBulk$ model,
primarily because the $\GRS$ model predicts a wider distribution in
$\abs{\Delta\eta}$. The rise in acceptance for the $\GRS$ model is
primarily due to the narrowing of the $\abs{\eta}$ distribution with
increasing resonance mass.

\begin{figure}[th!b]
\centering
\includegraphics[width=0.49\textwidth]{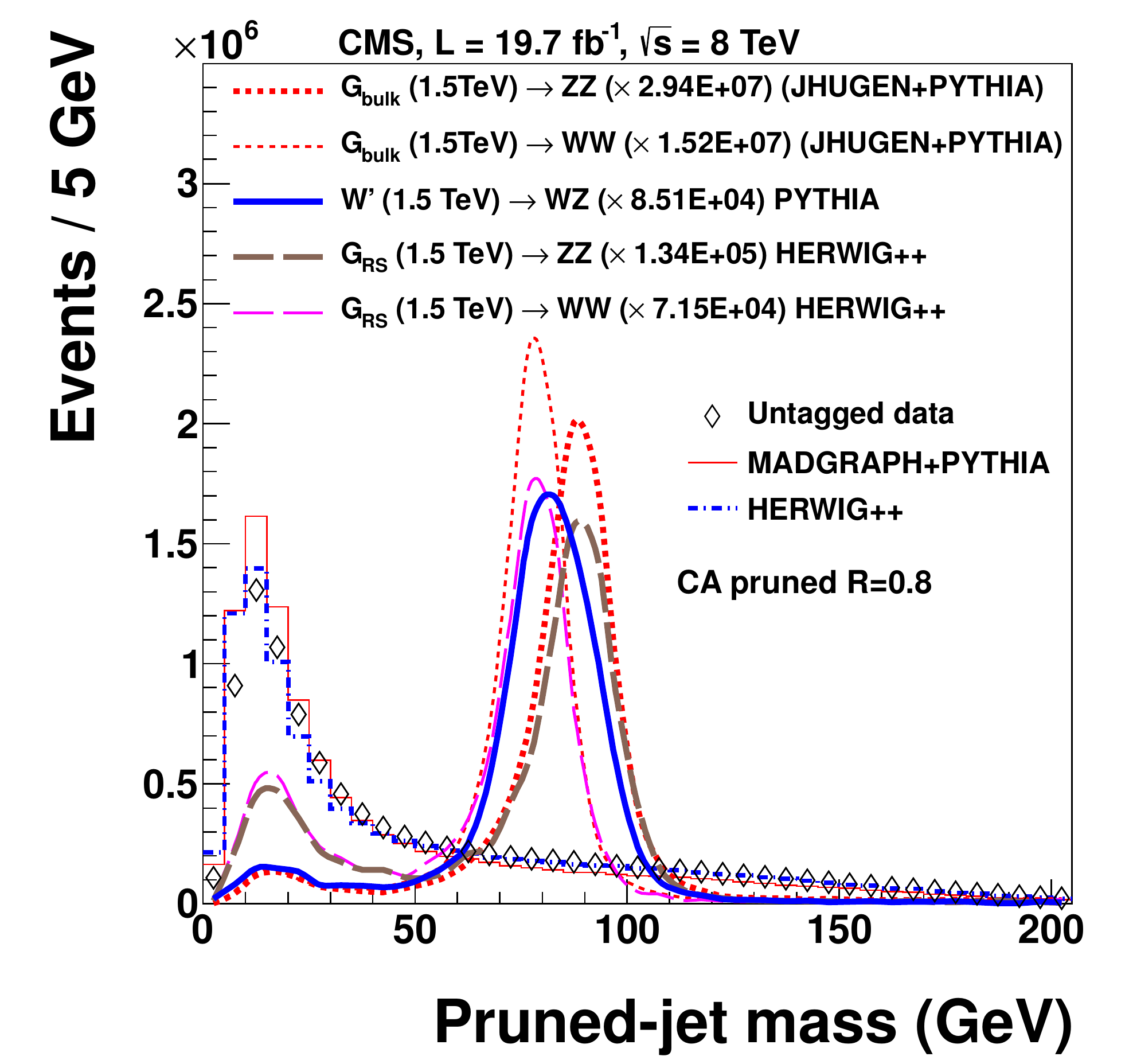}
\includegraphics[width=0.49\textwidth]{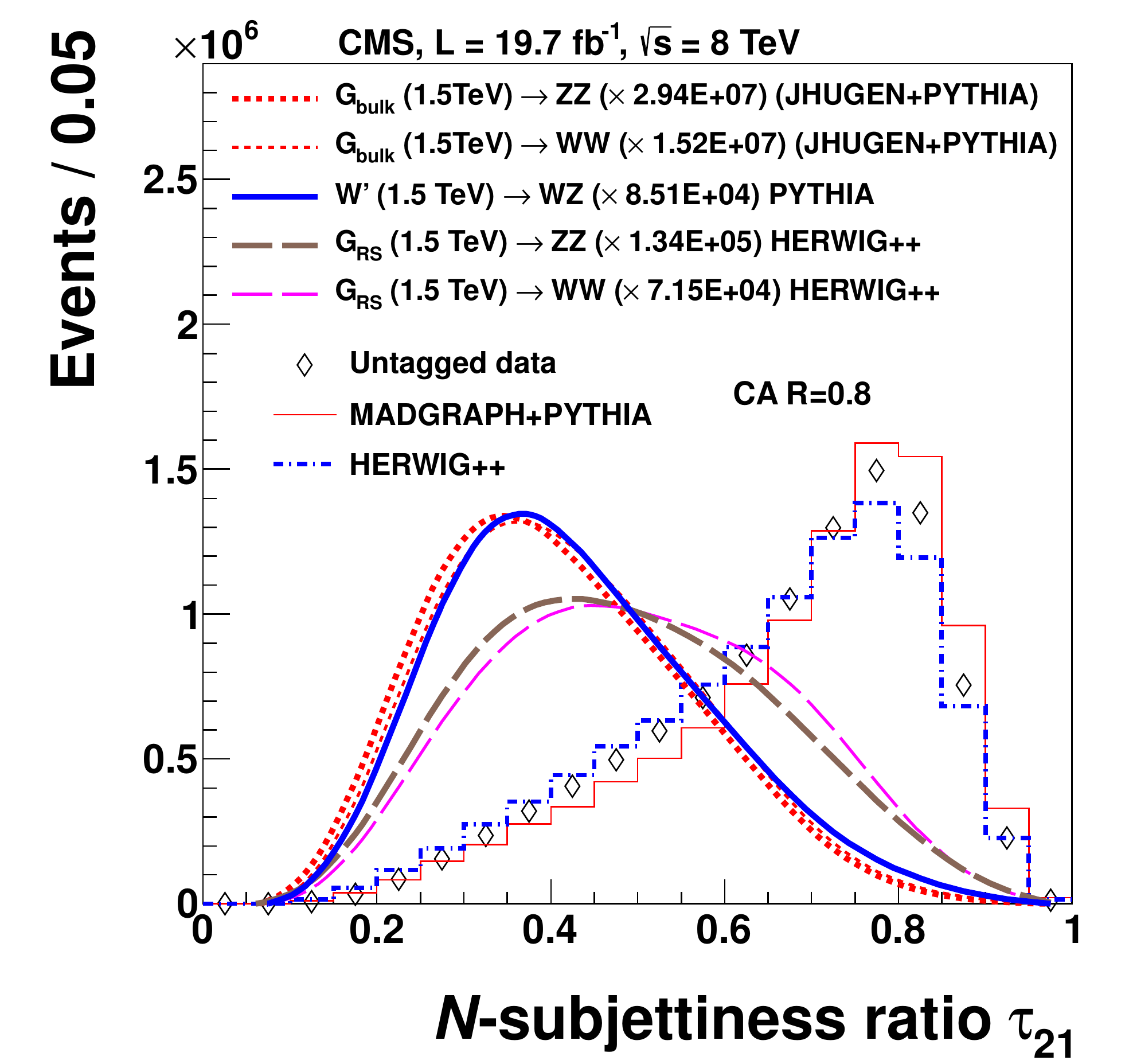}
  \caption{Distribution for (left) pruned-jet mass $m_\mathrm{j}$ and (right) jet
   $N$-subjettiness ratio $\tau_{21}$ in data, and in simulations
   of signal and background events. All simulated distributions are
   scaled to match the number of events in data. \MADGRAPH/\PYTHIA and
   \HERWIG{++} refer to QCD multijet event
   simulations.\label{fig:taggingvariables}}
\end{figure}

The two jets of highest $\pt$ are chosen as candidates of highly
boosted \PW\ or \cPZ\ bosons decaying to quarks, and passed through a
tagging algorithm based on jet
pruning~\cite{jetpruning1,jetpruning2,catop_cms,topwtag_pas}. Each jet
is reclustered using all the particles that form the original CA jet,
associating with each step of the recombination procedure a measure of
the jet's ``softness''. The CA clustering algorithm starts from a set
of ``protojets'' given by the PF particles. Iteratively these
protojets are combined with each other until a set of jets is
found. Given two protojets $i$, $j$ of transverse momenta $\pt^i$ and
$\pt^j$, the recombination, that is the sum of their transverse
momenta $\pt^p$, is considered soft if: (i) its hardness $z$ is found
to be $z<0.1$, where $z$ is the smaller of the two values of
$\pt^i/\pt^p$ and $\pt^j/\pt^p$, or (ii) when the two protojets have a
distance $\Delta R$ larger than some $D_{\text{cut}}$, where the value
of $D_{\text{cut}}$ is given by $m^\text{orig}/\pt^\text{orig}$, with
$m^\text{orig}$ and $\pt^\text{orig}$ representing the mass and $\pt$
of the original CA jet. If a recombination is identified as soft, the
protojet with smaller \pt is discarded. If the pruned jet has a mass
within $70 <m_\mathrm{j}<100\GeVcc$, it is tagged as a $\PW/\cPZ$
candidate. This mass requirement was optimized specifically for this
analysis.  The distributions of $m_\mathrm{j}$ for data, and for
simulated signal and background samples, are shown in
Fig.~\ref{fig:taggingvariables}~(left).  Fully merged jets from \PW\
and \cPZ\ decays are expected to generate a peak at $m_\mathrm{j}
\approx 80$--90\GeVcc, while jets from multijet events and
not-fully-merged \PW\ and \cPZ\ bosons give rise to a peak around
20\GeV. The disagreement observed at small values of
$m_\mathrm{j}$~\cite{CMSjetmass} can be ignored, as the \PW\ and \cPZ\
candidates with $m_\mathrm{j}<70$\GeVcc\ are not considered in the
analysis and the overall background normalization is determined with a
fit to the data.

We achieve additional discrimination against multijet events by
considering the distribution of jet constituents relative to the jet
axis. In particular, we quantify how well the constituents of a given
jet can be arranged into $N$ subjets. This is done by reconstructing
the full set of jet constituents (before pruning) with the \kt
algorithm~\cite{ktalg} and halting the reclustering when $N$
distinguishable protojets are formed.  
The directions of the $N$ jets are used as the reference axes to
compute the
$N$-subjettiness~\cite{Thaler:2010tr,Thaler:2011gf,Stewart:2010tn}
$\tau_{N}$ of the original jet, defined as
\begin{equation}
\label{eq:subjettiness}
\tau_N = \frac{1}{d_{0}} \sum_{k} \ptk\,\min( \Delta R_{1,k}, \Delta R_{2,k},\ldots,\Delta R_{N,k}),
\end{equation}
where $\ptk$ is the $\pt$ of the particle constituent $k$ of the
original jet, and $\Delta R_{n,k}$ is its angular distance from the
axis of the $n$th subjet (with $n=1, 2,\ldots,N$). The normalization
factor $d_{0}$ for $\tau_N$ is $d_{0}= \sum_{k} \ptk R_{0}$, with
$R_{0}$ set to the distance parameter $R$ of the original CA jet. To
improve the discriminating power, we perform a one-pass optimization
of the directions of the subjet axes by minimizing
$\tau_{N}$~\cite{Thaler:2011gf,JME-13-006}.  By using the smallest
$\Delta R_{n,k}$ to weight the value of $\ptk$ in
Eq.~(\ref{eq:subjettiness}), $\tau_N$ yields small values when the jet
originates from the hadronization of $N$ quarks. We therefore use the
ratio $\tau_{21}=\tau_{2} / \tau_{1}$ as a discriminant between the
two-pronged $\PW\to\cPaq \Pq'$ or $\cPZ\to\cPaq \Pq$ decays and single
jets in multijet events. The discriminating power of $\tau_{21}$ for
different resonance models can be seen in
Fig.~\ref{fig:taggingvariables} (right). The MC simulations of
multijet background and the data peak near $\approx$0.8, whereas the
signal distributions have a larger fraction of events at smaller
values of $\tau_{21}$. We found a slightly better significance using
N-subjettiness without pruning, taking pileup uncertainties into
account.

Differences are observed in signal distributions predicted with
\HERWIG{++} (for $\GRS$), with \PYTHIA ($\Pq^*$, $\PWpr$), and with
{\sc jhugen}/\PYTHIA ($\GBulk$), for the mass $m_\mathrm{j}$ of pruned
jets and for $\tau_{21}$. These differences arise from unlike
polarization of the vector bosons in the various signals models and from
differences between \HERWIG{++} and \PYTHIA in the modelling of
the showering and hadronization of
partons. In particular, values for the polarization of the vector
bosons are related to different predictions for $\tau_{21}$ in the
$\GRS$ and $\GBulk$ models as noted in Ref.~\cite{JME-13-006}.
Differences in the modelling of the small $m_\mathrm{j}$ regions for
pruned jets have been observed previously~\cite{CMSjetmass}.  The
showering and hadronization differences are taken into account in the
estimation of systematic uncertainties, as discussed below.

We select ``high-purity'' (HP) $\PW/\cPZ$ jets by requiring $\tau_{21}
\leq 0.5$, and ``low-purity'' (LP) $\PW/\cPZ$ jets by requiring $ 0.5
< \tau_{21} < 0.75$. Events with just one $\PW/\cPZ$ tag are
classified according to these two categories. The events with two
$\PW/\cPZ$-tagged jets are always required to have one HP $\PW/\cPZ$
tag, and are similarly divided into HP and LP events, depending on
whether the other $\PW/\cPZ$-tagged jet is of high or low purity. The
selection criterion for the HP category is chosen to give optimal
average performance for the models used in this search. The LP
category adds sensitivity, especially at large values of
$m_\mathrm{jj}$, where the rate in the HP category drops along with
the background rate.

\begin{figure}[th!b]
\centering
\includegraphics[width=0.49\textwidth]{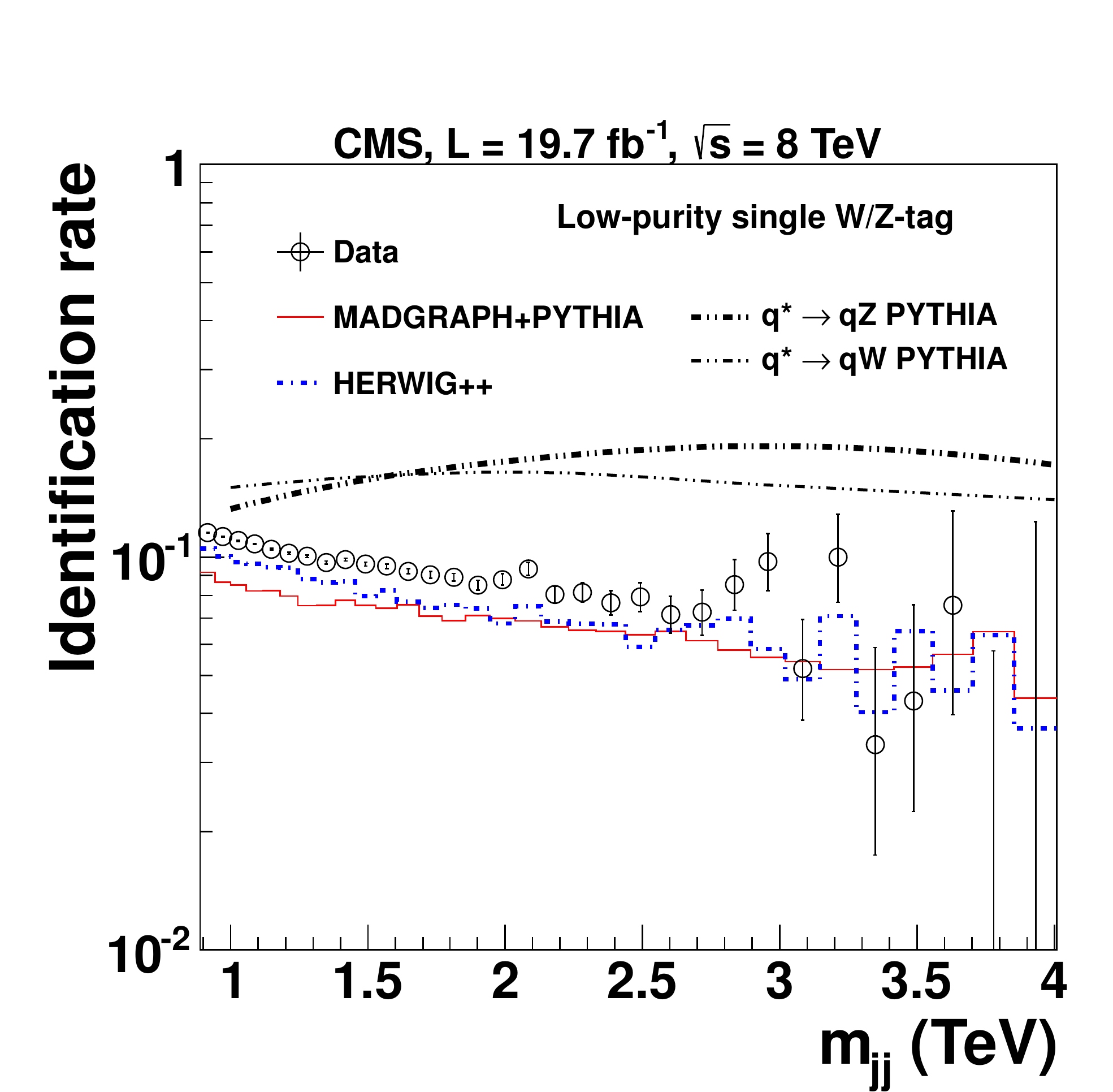}
\includegraphics[width=0.49\textwidth]{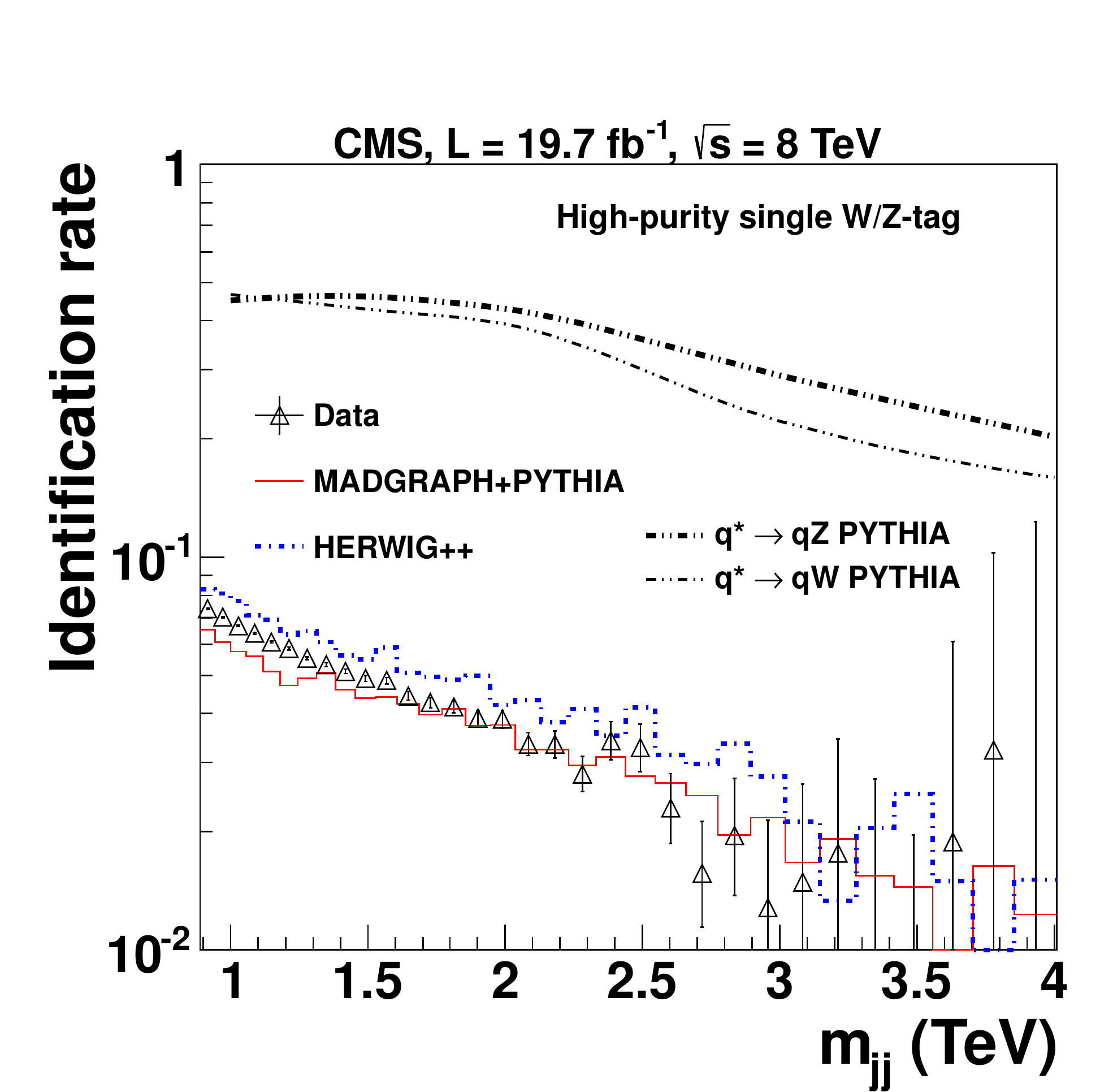}
\includegraphics[width=0.49\textwidth]{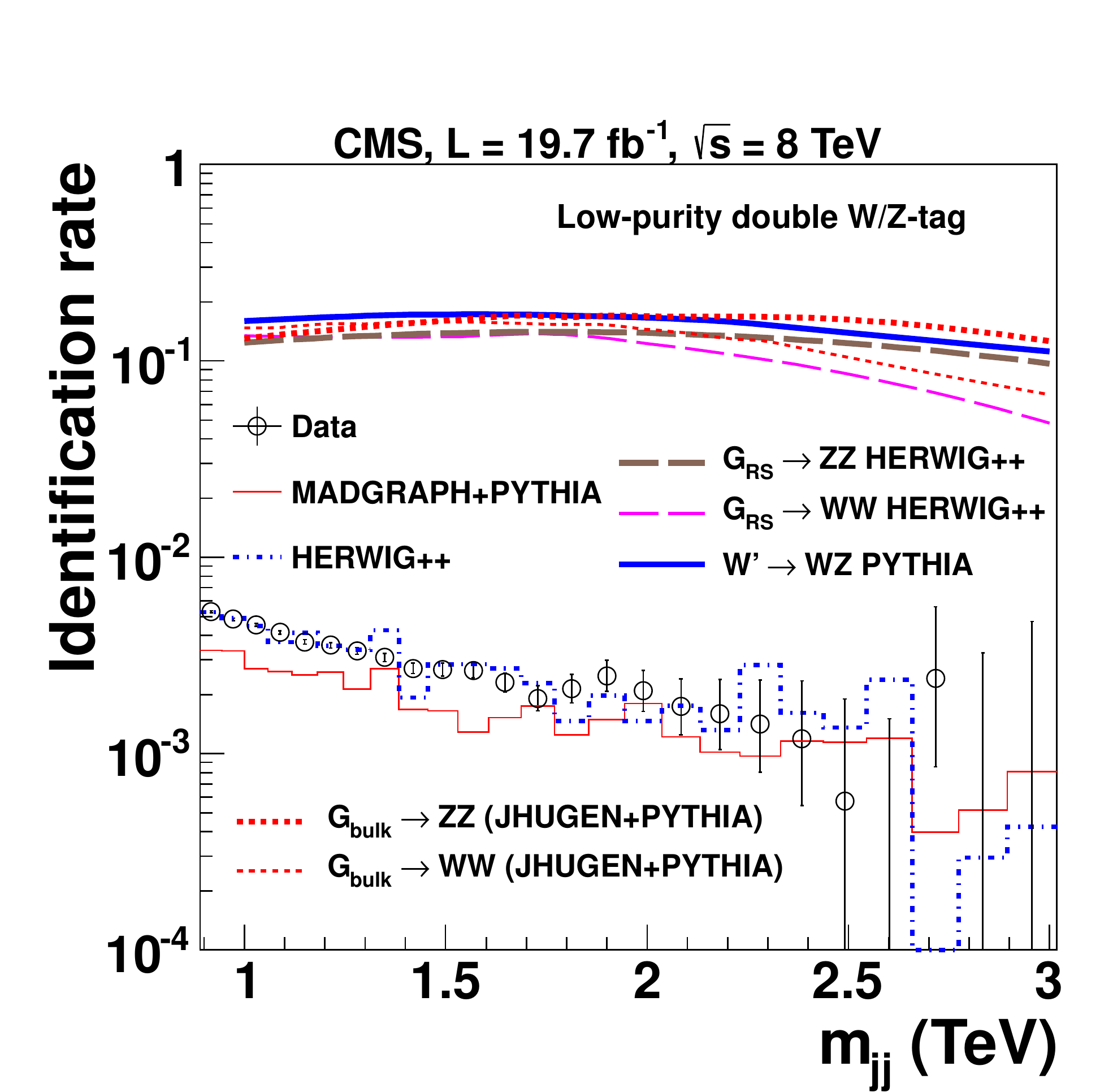}
\includegraphics[width=0.49\textwidth]{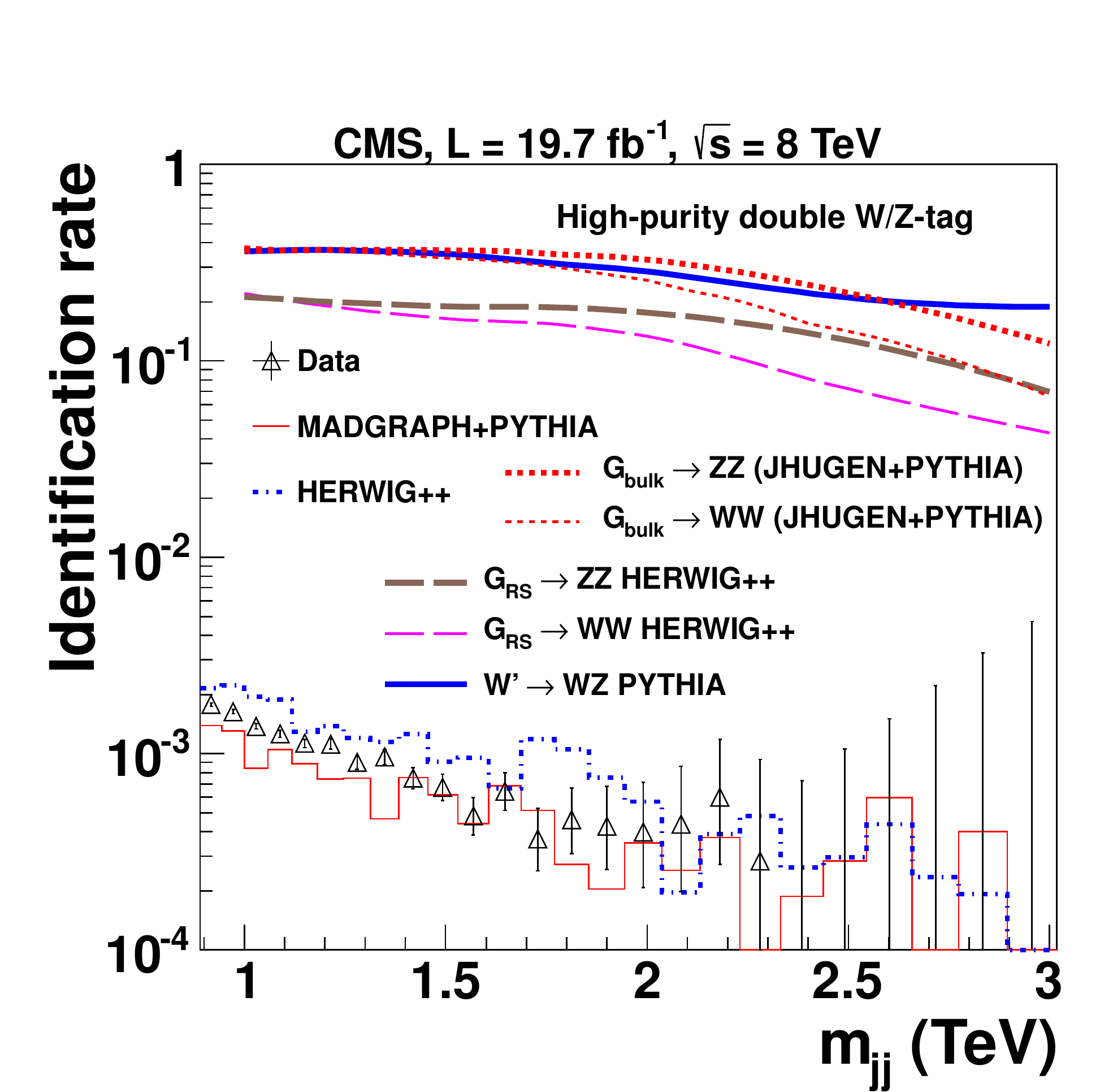}

\caption{Identification rate for \PW\ and \cPZ\ boson selections
  as a function of $m_\mathrm{jj}$ for quark and gluon
  jets in data and in simulation of background events,
  and for jets from \PW\ and \cPZ\ bosons in simulation of signal events,
  with (upper left) one LP or (upper
  right) HP $\PW/\cPZ$-tag, and the fraction of (lower left)
  doubly-tagged events in the LP and (lower right) HP category. The
  identification rate is computed for $\PW/\cPZ\to\cPaq \Pq' \to \text{jets}$
  events, where the jets have $\abs{\eta} < 2.5$ and
  $\abs{\Delta\eta}<1.3$.  \MADGRAPH/\PYTHIA and \HERWIG{++} refer to
  QCD multijet event simulations.\label{fig:WZefficiencies}}
\end{figure}

The identification rates expected for the \PW\ and \cPZ\ selection
criteria for signal and background events in different event
categories are shown in Fig.~\ref{fig:WZefficiencies} as a function of
$m_\mathrm{jj}$. As expected from Fig.~\ref{fig:taggingvariables}
(right), the background simulation shows disagreements in modelling
the identification rate for background events in data; however, the
dependence as a function of \pt is well modelled.  While the
background simulation is not used to model the background in the
analysis, it shows how well the simulation models the \pt dependence
of substructure variables. The $\PW/\cPZ$-tagging efficiency for
signal events in the HP categories drops at high \pt, while it is more
stable in the LP categories, primarily because the $\tau_{21}$
distribution is \pt-dependent.

The modelling of the signal efficiency is cross-checked through a
$\PW$-tagging efficiency estimated using merged $\PW\to\cPaq \Pq'$
decays in $\ttbar$ events~\cite{JME-13-006}. The efficiency is
obtained using $\ell$+jets events with two b-tagged jets, one of which
has $\pt > 200$\GeVc. Such events are dominated by $\ttbar$
production. The data are compared to simulated $\ttbar$ events,
generated with {\MADGRAPH}, interfaced to \PYTHIA for parton
showering, and provide scale factors of \scalefactorHP~and
\scalefactorLP, respectively, for HP and LP events.  These values are
derived following the method described in Ref.~\cite{JME-13-006} for
the selections applied in this analysis, and are used to match the
simulated samples to data. The uncertainties in the scale factors
contribute to the systematic uncertainty in the selection efficiency
for signal.

\begin{figure}[th!b]
\centering
\includegraphics[width=0.49\textwidth]{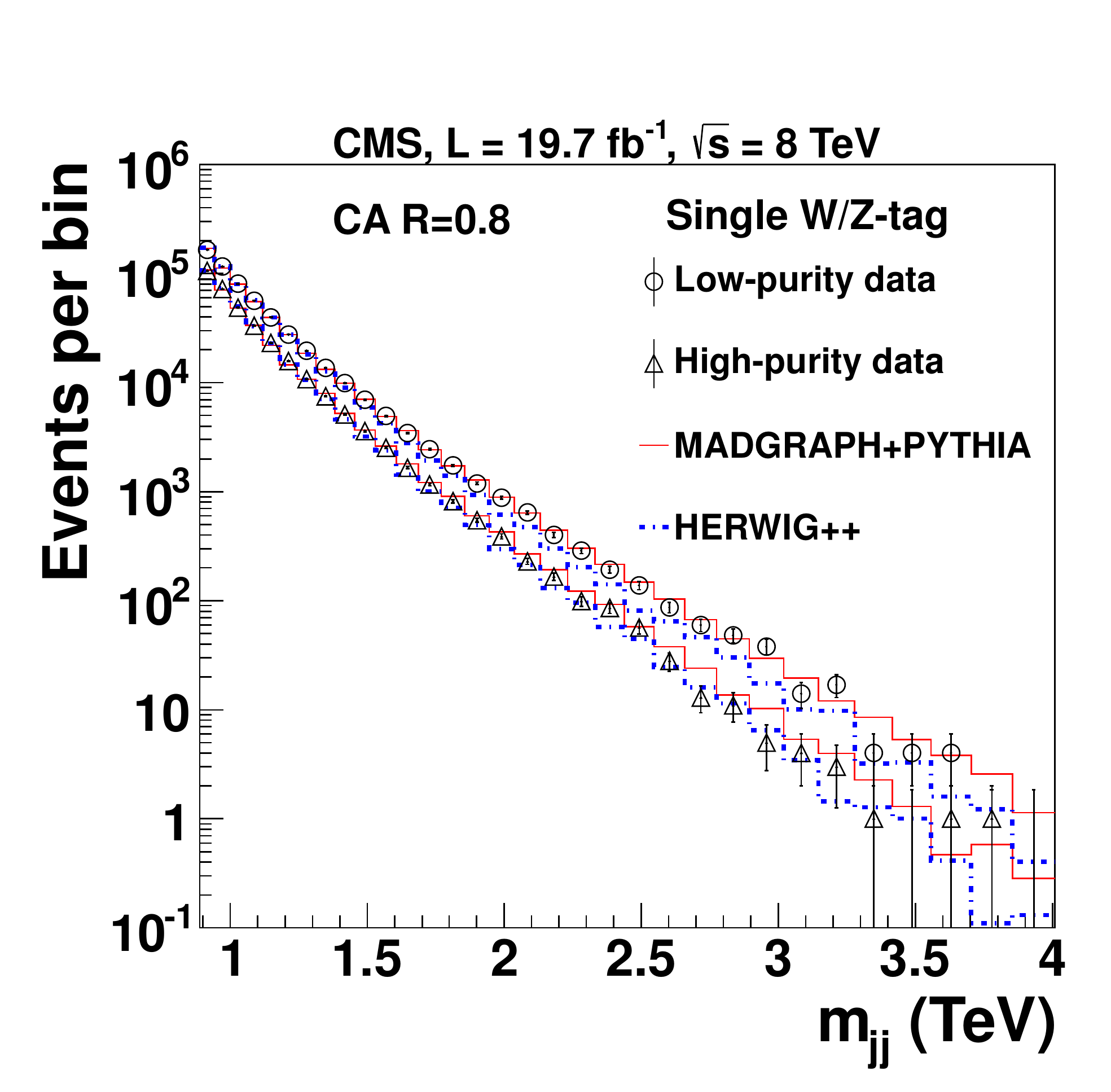}
\includegraphics[width=0.49\textwidth]{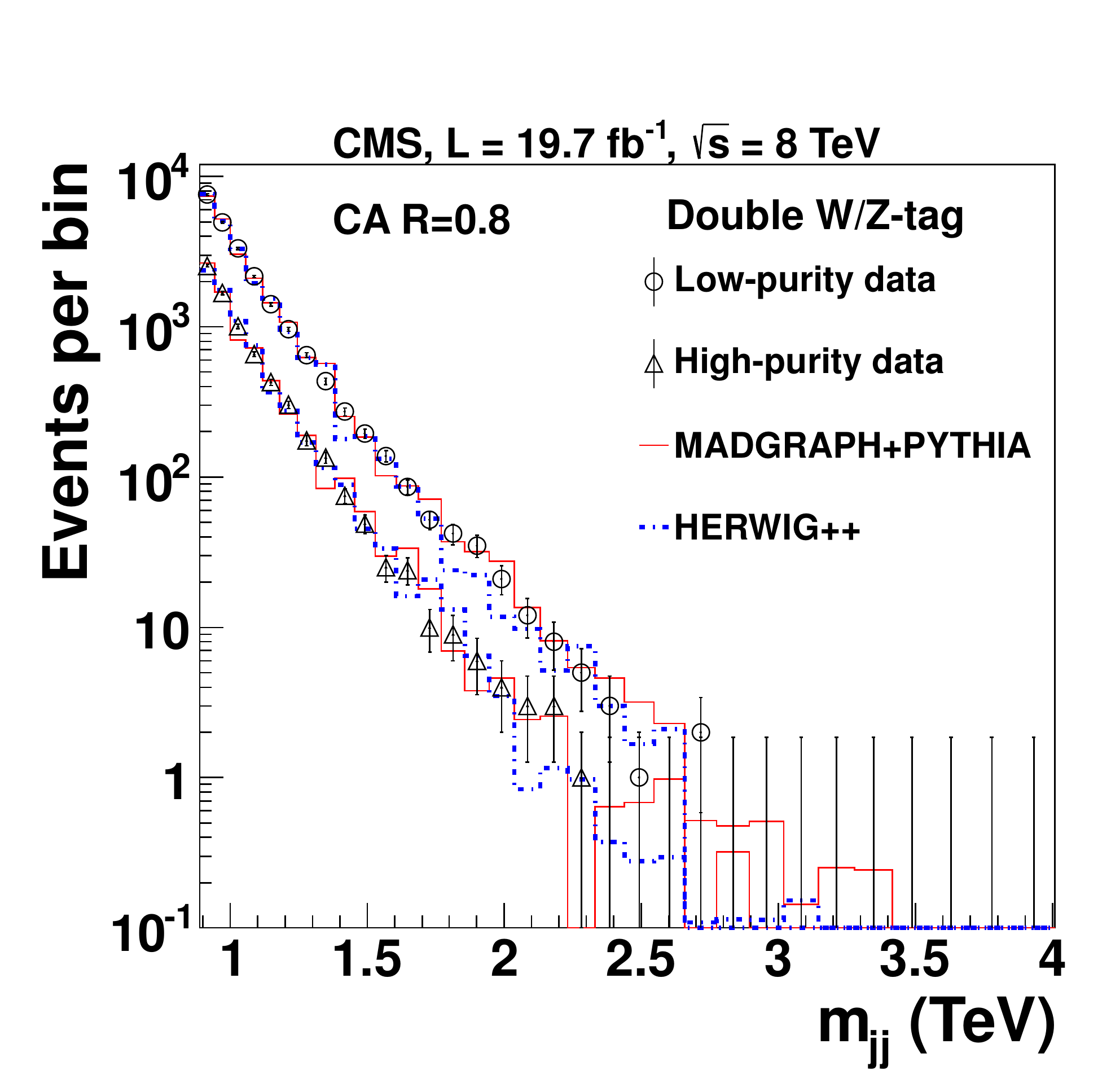}

 \caption{The $m_\mathrm{jj}$ distributions for (left) singly and
  (right) doubly tagged events in data, and for QCD multijet
  (\MADGRAPH/\PYTHIA and \HERWIG{++}) simulations, normalized to
  data.\label{fig:mjjqV}}
\end{figure}

The $m_\mathrm{jj}$ distributions for singly and doubly tagged LP and
HP event samples are shown in Fig.~\ref{fig:mjjqV}. These
distributions provide the basis for the search. The analogous
distributions from \MADGRAPH/\PYTHIA and \HERWIG{++} multijet simulations,
normalized to the number of events in data, are shown. Only the
dominant background from multijet production without systematic
uncertainties is shown in this comparison. The prediction from
\HERWIG{++} decreases more steeply with an increase in $m_\mathrm{jj}$
than that for \MADGRAPH/\PYTHIA. We estimate from simulation that
backgrounds from $\ttbar$, W+jets and Z+jets events with the vector
bosons decaying into quark final states contribute less than 2\% of
total background.

\section{The search for a peak in the mass spectrum}
\label{sec:background}

\begin{figure}[htb]
\centering
\includegraphics[width=0.49\textwidth]{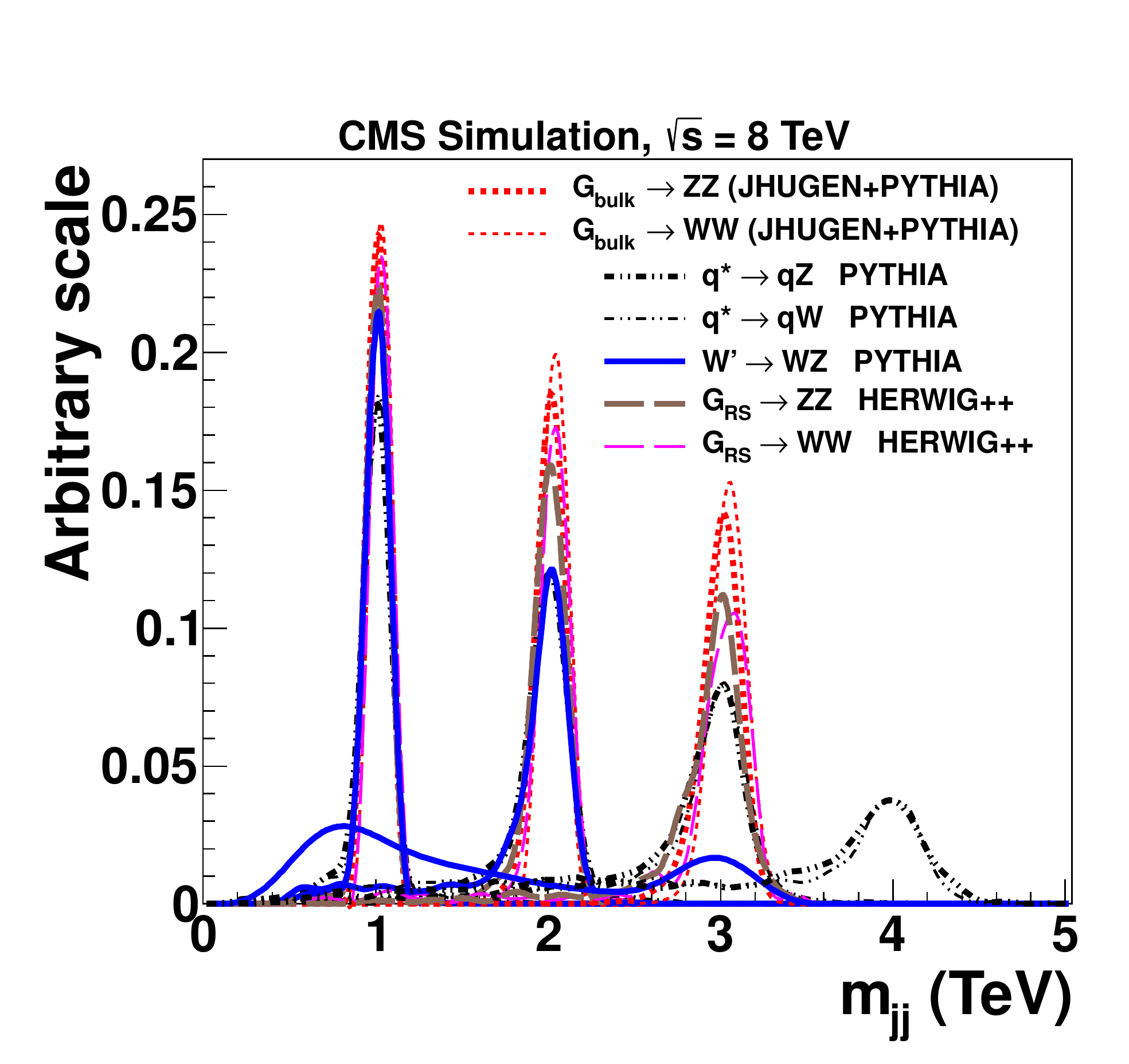}
\caption{Distribution in $m_\mathrm{jj}$ expected in the HP categories
  corresponding to resonance masses of 1, 2, 3\TeV, for all models,
  and 4 \TeV, for $\Pq^*$ models. All distributions are normalized to
  the same area. \label{fig:highresonanceshape}}
\end{figure}

Figure~\ref{fig:highresonanceshape} shows the $m_\mathrm{jj}$
distributions expected for the HP category of $\GRS \to
\cPZ\cPZ/\PW\PW$, $\GBulk\to\cPZ\cPZ/\PW\PW$, $\PWpr \to \PW\cPZ$, and
$\Pq^*\to\Pq\PW/\Pq\cPZ$, for four resonance masses. A linear
interpolation between a set of reference distributions (corresponding
to masses of 1.0, 1.5, 1.8, 2.0, 2.2, 2.5, 3.0, and 4.0\TeV) is used
to obtain the expected distribution for other values of resonance
mass. Because of the interplay between the PDF and the resonance
width, the $\PWpr$ distribution for large resonance masses is also
characterized by a contribution at small masses that peaks near
$\approx$0.8\TeV. This search is not sensitive to this component
because of the overwhelming background from multijet production.
This feature is not observed for the other signal models, which
assume a narrow width.

Background from multijet events is modelled by a smoothly falling
distribution for each event category, given by the empirical
probability density function
\begin{equation}
P_D(m_\mathrm{jj}) = \frac{P_{0} (1 - m_\mathrm{jj}/\sqrt{s})^{P_{1}}}{(m_\mathrm{jj}/\sqrt{s})^{P_{2}}} \ .
\label{eqParam}
\end{equation}
\noindent For each category, the normalization factor $P_0$ and the
two parameters $P_1$ and $P_2$ are treated as uncorrelated. This
parameterization was deployed successfully in searches in dijet mass
spectra~\cite{cmsdijet}. A Fisher F-test~\cite{Ftest} is used to check
that no additional parameters are needed to model the individual
background distribution, for each of the four cases considered.

We search for a peak on top of the falling background spectrum by
means of a maximum likelihood fit to the data. The likelihood $\mathcal{L}$, computed using events binned as a function of $m_\mathrm{jj}$,
is written as
\begin{equation} \mathcal{L} = \prod_{i}
  \frac{\lambda_{i}^{n_{i}}\re^{-\lambda_{i}}}{n_{i}!},
\end{equation}
where ${\lambda_{i}} = {\mu}{N_{i}(S)} + {N_{i}(B)}$,
$\mu$ is a scale factor for the signal, $N_i(S)$ is the number
expected from the signal, and $N_i(B)$ is the number expected from
multijet background. The parameter $n_i$ quantifies the number of
events in the $i^\mathrm{th}$ $m_\mathrm{jj}$ mass bin.
The background $N_i(B)$ is described by the functional form of
Eq.~(\ref{eqParam}). While maximizing the likelihood as a function of
the resonance mass, $\mu$ as well as the parameters of the background
function are left floating.

\begin{figure}[th!b]
\centering

\includegraphics[width=0.49\textwidth]{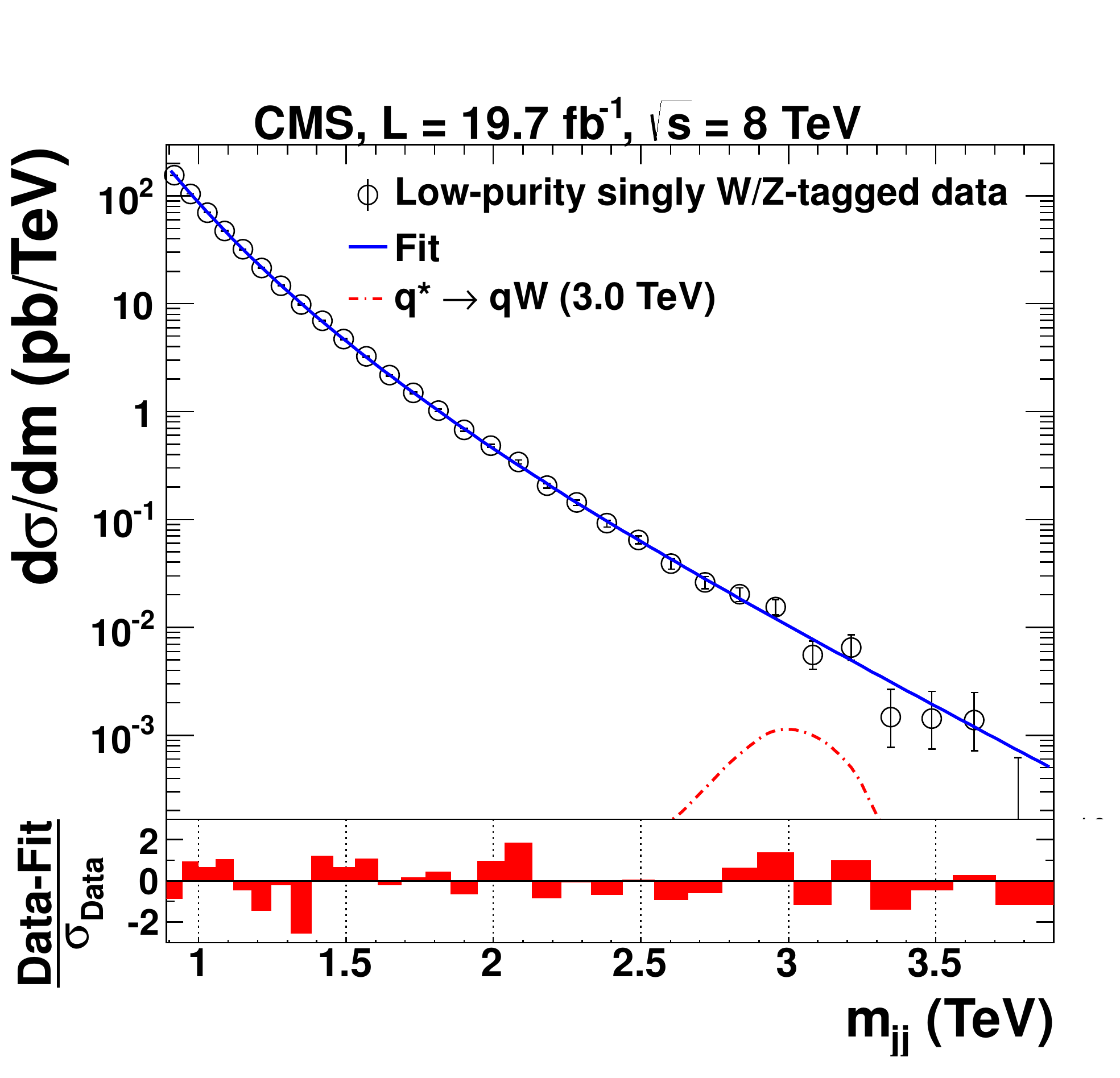}
\includegraphics[width=0.49\textwidth]{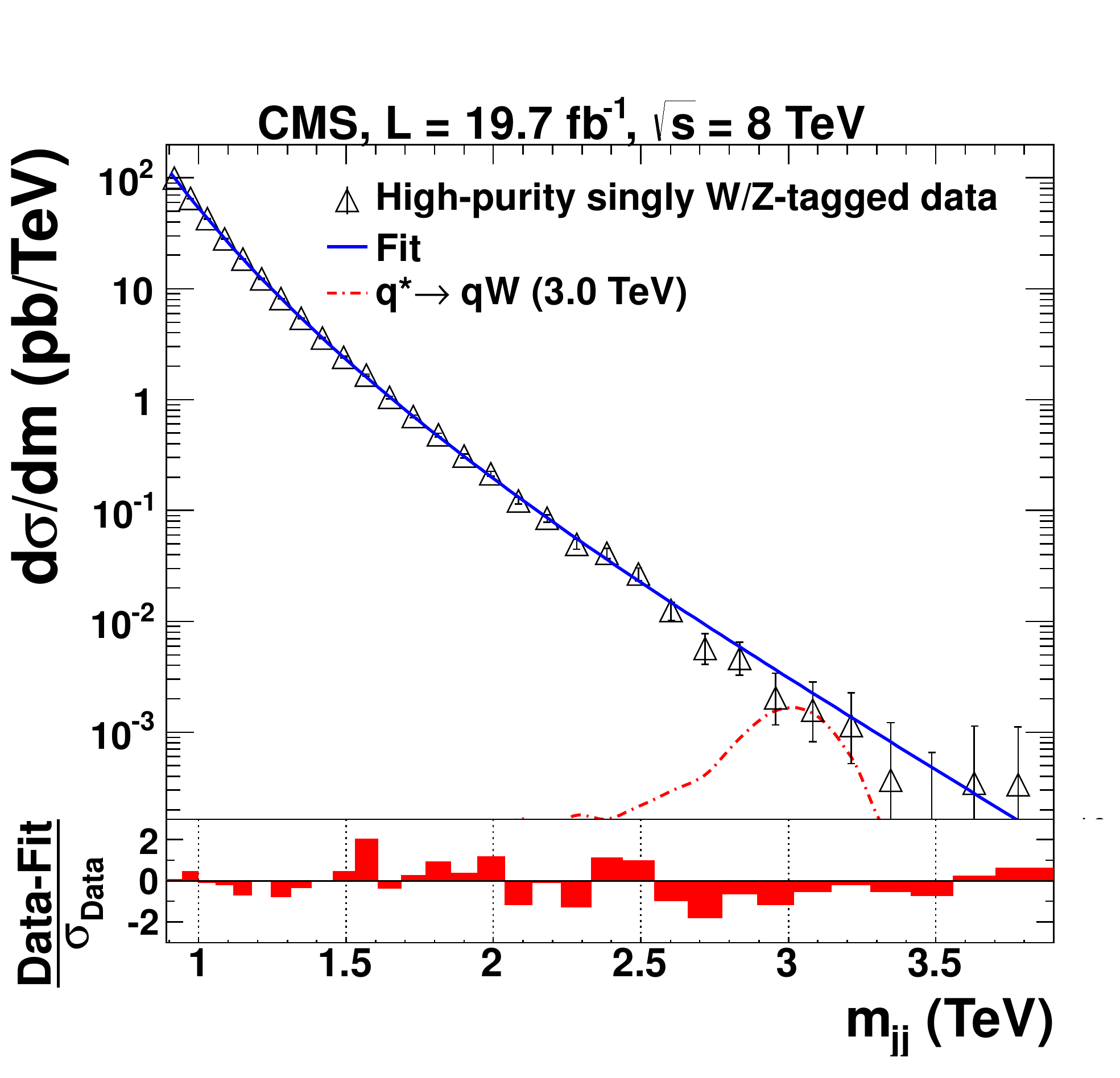}
\includegraphics[width=0.49\textwidth]{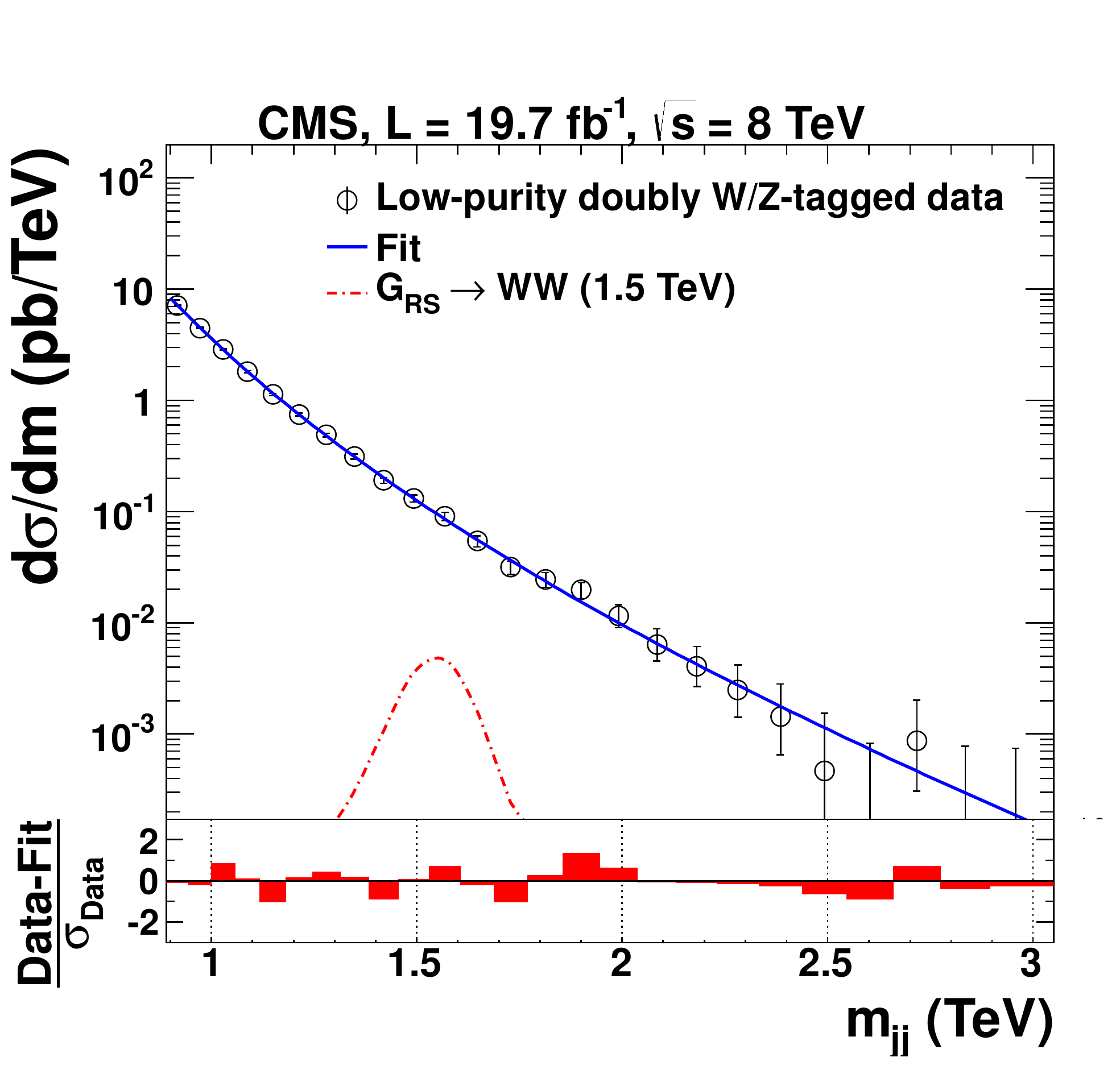}
\includegraphics[width=0.49\textwidth]{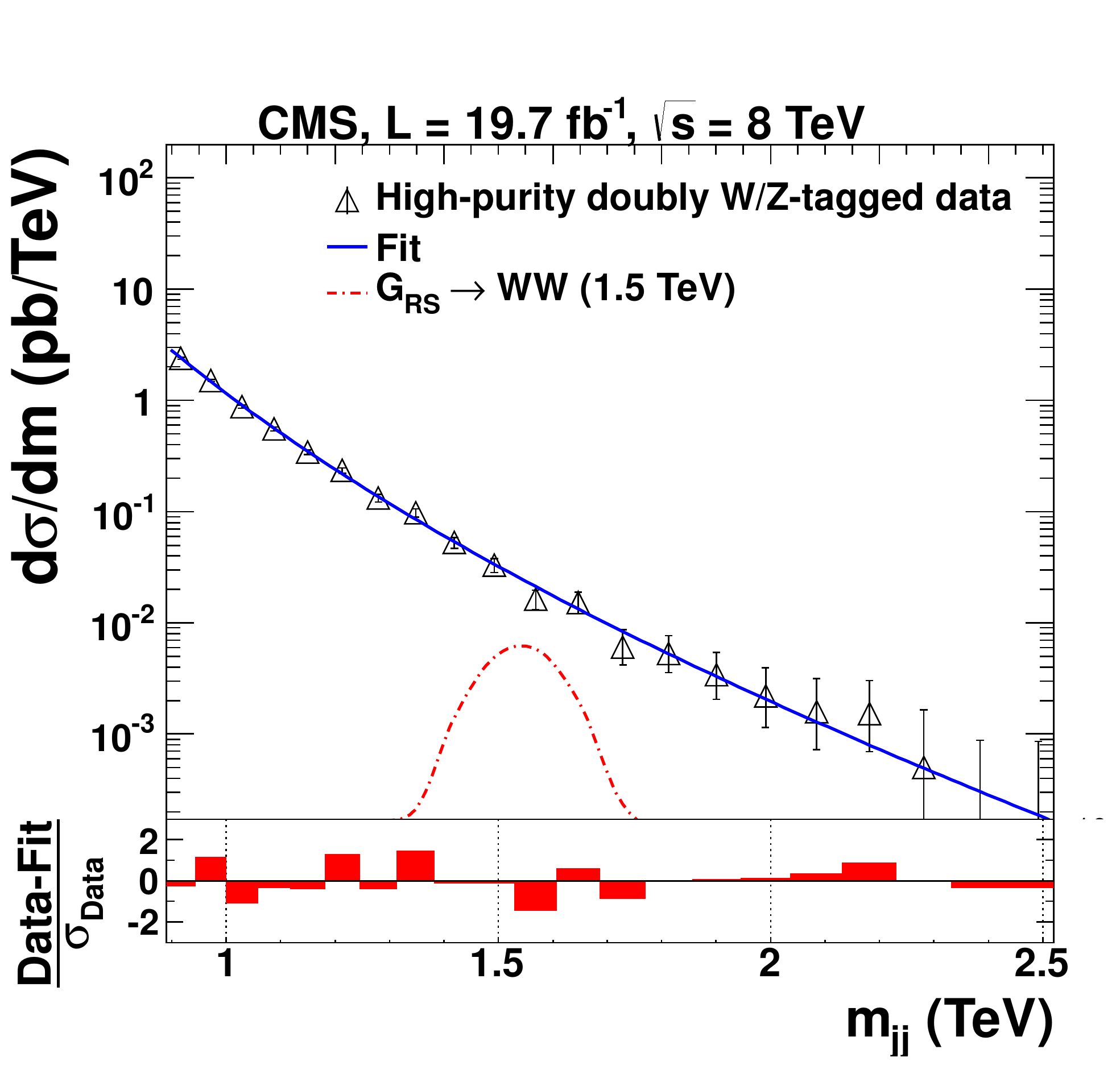}
 \caption{Distribution in $m_\mathrm{jj}$, respectively, for
   (upper left) singly-tagged LP events and (upper right) HP events, and for (lower left) doubly-tagged
   LP events and (lower right) HP events. The solid curves represent the
   results of fitting Eq.~(\ref{eqParam}) to the data. The
   distribution for $\Pq^*\to\Pq\PW$ and $\GRS \to\PW\PW$
   contributions, scaled to their corresponding cross sections, are
   given by the dash-dotted curves. The corresponding pull
   distributions
   ($\frac{\text{Data}-\text{Fit}}{\sigma_{\text{Data}}}$, where
   $\sigma_{\text{Data}}$ represents the statistical uncertainty in
   the data in a bin in $m_\mathrm{jj}$) are shown below each
   $m_\mathrm{jj}$ plot.\label{fig:BG}}
\end{figure}

Figure~\ref{fig:BG} shows the $m_\mathrm{jj}$ spectra in data with a
single $\PW/\cPZ$-tag, and with a double $\PW/\cPZ$-tag. The solid
curves represent the results of the maximum likelihood fit to the data,
fixing the number of expected signal events to 0, while the bottom
panels show the corresponding pull distributions, quantifying the
agreement between the background-only hypothesis and the data.
The expected contributions from $\Pq^*$ and $\GRS$
resonances for respective masses of 3.0 and 1.5\TeV, scaled to
their corresponding cross sections, are given by the dash-dotted
curves.

We quantify the consistency of the data with the null hypothesis as a
function of resonance mass for the benchmark models through the local
p-value. The largest local significance in the singly
$\PW/\cPZ$-tagged sample is observed for the hypothesis of a $\Pq^*
\to \Pq\PW$ resonance of mass 1.5\TeV, and is equivalent to an excess
of 1.8 standard deviations. The largest local significance in the
doubly tagged event sample corresponds to an excess of 1.3 standard
deviations for a $\GRS\to\PW\PW$ resonance of mass 1.9\TeV. Using the
$\GBulk\to\PW\PW/\cPZ\cPZ$ model, where the LP and HP categories
contribute in different proportions compared to the case for the
$\GRS\to\PW\PW$ model, yields no excess larger than one standard
deviation. 

Using pseudo-experiments, we estimated the probability of observing a
local statistical fluctuation of at least two standard deviations in
any mass bin. This probability corresponds to an equivalent global
significance of one standard deviation.  The $m_\mathrm{jj}$ distributions
are used to set upper limits on the product of the production cross
sections and decay branching fractions for the benchmark models.

\section{Systematic uncertainties}
\label{sec:systematics}

The largest contributions to systematic uncertainties are associated
with the modelling of the signal, namely the determination of the
$\PW/\cPZ$-tagging efficiency, jet energy scale (JES), jet energy
resolution (JER), and integrated luminosity.

The uncertainty in the efficiency for singly $\PW/\cPZ$-tagged events
is estimated using the $\ell$+jets control sample from $\ttbar$ events
described above. Uncertainties of \scalefactorHPu~and
\scalefactorLPu~in the respective scale factors for HP and LP tagging
include contributions from control-sample statistical uncertainties,
and the uncertainties in the JES and JER for pruned jets. Since the
scale factors are estimated only in the kinematic regime of the
$\ttbar$ sample, where the \PW\ decay products merge and the \cPqb\
quarks are reconstructed as separate jets, we use the simulation just
to extrapolate to larger $\PW/\cPZ$-jet \pt. The efficiency is
therefore estimated as a function of \pt for two showering and
hadronization models, using $\GBulk$ samples generated with the {\sc
  jhugen} event generator interfaced to \PYTHIA and \HERWIG{++}. The
differences are respectively within 4\% and 12\% for HP and LP
tagged jets, significantly smaller than the statistical uncertainties
in the scale factors. Other systematic uncertainties in tagging
efficiency are even smaller. Because of the rejection of charged
particles not originating from the primary vertex, and the application
of pruning, the dependence of the $\PW/\cPZ$-tagging efficiency on
pileup is weak, and the uncertainty in the modelling of the pileup
distribution is $<$1.5\%. These systematic contributions refer to a
singly $\PW/\cPZ$-tagged jet, and are applied to each of the two
leading jets in doubly $\PW/\cPZ$-tagged events.

The JES has an uncertainty of
1--2\%~\cite{JME-JINST,Collaboration:2013dp}, and its \pt and $\eta$
dependence is propagated to the reconstructed value of
$m_\mathrm{jj}$, yielding an uncertainty of 1\%, regardless
of the resonance mass. The impact of this uncertainty on the
calculated limits is estimated by changing the dijet mass in the
analysis within its uncertainty. The JER is known to a precision of
10\%, and its non-Gaussian features observed in data are well
described by the CMS simulation~\cite{JME-JINST}. The effect of the
JER uncertainty in the limits is also estimated by changing the
reconstructed resonance width within its uncertainty. The integrated
luminosity has an uncertainty of 2.6\%~\cite{LUM-13-001}, which is
also taken into account in the analysis. The uncertainty related to
the PDF used to model the signal acceptance is estimated from the
eigenvectors of the CTEQ66~\cite{cteq} and MRST2006~\cite{mrst2006}
sets of PDF. The envelope of the upward and downward variations of the
estimated acceptance for the two sets is assigned as uncertainty and
found to be 5\% -- 15\% in the resonance mass range of interest. A
summary of all systematic uncertainties is given in
Table~\ref{table:uncertainties}.

\begin{table}[]
\begin{center}
  \topcaption{Summary of systematic uncertainties.  The labels HP and
    LP refer to high-purity and low-purity event categories,
    respectively.}
\begin{tabular}{ llll }
\hline
Source         &  Relevant quantity         & LP uncertainty (\%)  & HP uncertainty (\%)   \\
\hline
Jet energy scale       & Resonance shape    & 1  & 1    \\
Jet energy resolution    & Resonance shape    & 10 & 10     \\
W-tagging    & Efficiency (per jet)  & \scalefactorHPuNoPer & \scalefactorLPuNoPer  \\
Tagging $\pt$-dependence  & Efficiency (per jet)  & $<$4 & $<$12 \\
Pileup	           & Efficiency (per jet)    & $<$1.5 & $<$1.5    \\
Integrated luminosity    & Yield (per event)    & 2.6  & 2.6  \\
PDF       & Yield (per event)    & 5--15  & 5--15  \\
\hline
\end{tabular}
\label{table:uncertainties}
\end{center}
\end{table}

\section{Results}
\label{sec:results}

The asymptotic approximation~\cite{AsymptCLs} of the LHC
$\mathrm{CL_s}$ method~\cite{CLs1,CLs3} is used to set upper limits on
the cross sections for resonance production. The dominant sources of
systematic uncertainties are treated as nuisance parameters associated
with log-normal priors in those variables, following the methodology
described in Ref.~\cite{ATLASCMSstat}. For a given value of the
signal cross section, the nuisance parameters are fixed to the values
that maximize the likelihood, a method referred to as
profiling. The dependence of the likelihood on parameters used to
describe the background in Eq.~(\ref{eqParam}) is removed in the same
manner, and no additional systematic uncertainty is therefore assigned
to the parameterization of the background.

The HP and LP event categories are combined into a common likelihood,
with the two uncertainties in the $\PW/\cPZ$-tagging efficiencies
considered to be anticorrelated between HP and LP tagging because of
the exclusive selection on $\tau_{21}$, while the remaining systematic
uncertainties in signal are taken as fully correlated. The variables
describing the background uncertainties are treated as uncorrelated
between the two categories. The LP category contributes to the
sensitivity of the analysis, especially at large values of
$m_\mathrm{jj}$. The combined expected limits on the $\GRS \to\PW\PW$
production cross sections are, respectively, a factor of 1.1 and 1.6
smaller at $m_\mathrm{jj}=1.0$\TeVcc and 2.9\TeVcc than the limit
obtained from the HP category alone.

\begin{figure*}[h!tb]
\centering
\includegraphics[width=0.49\textwidth]{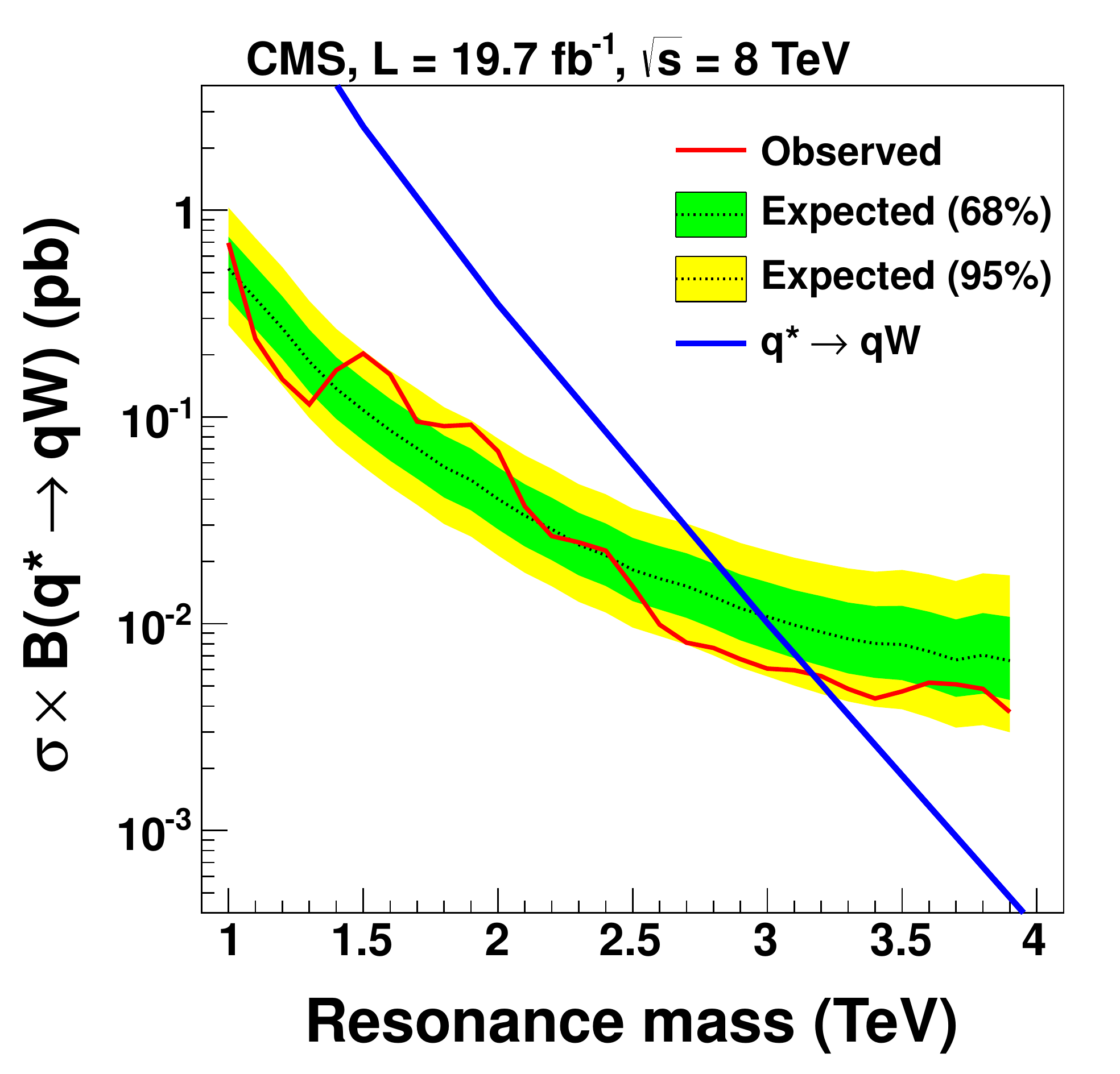}
\includegraphics[width=0.49\textwidth]{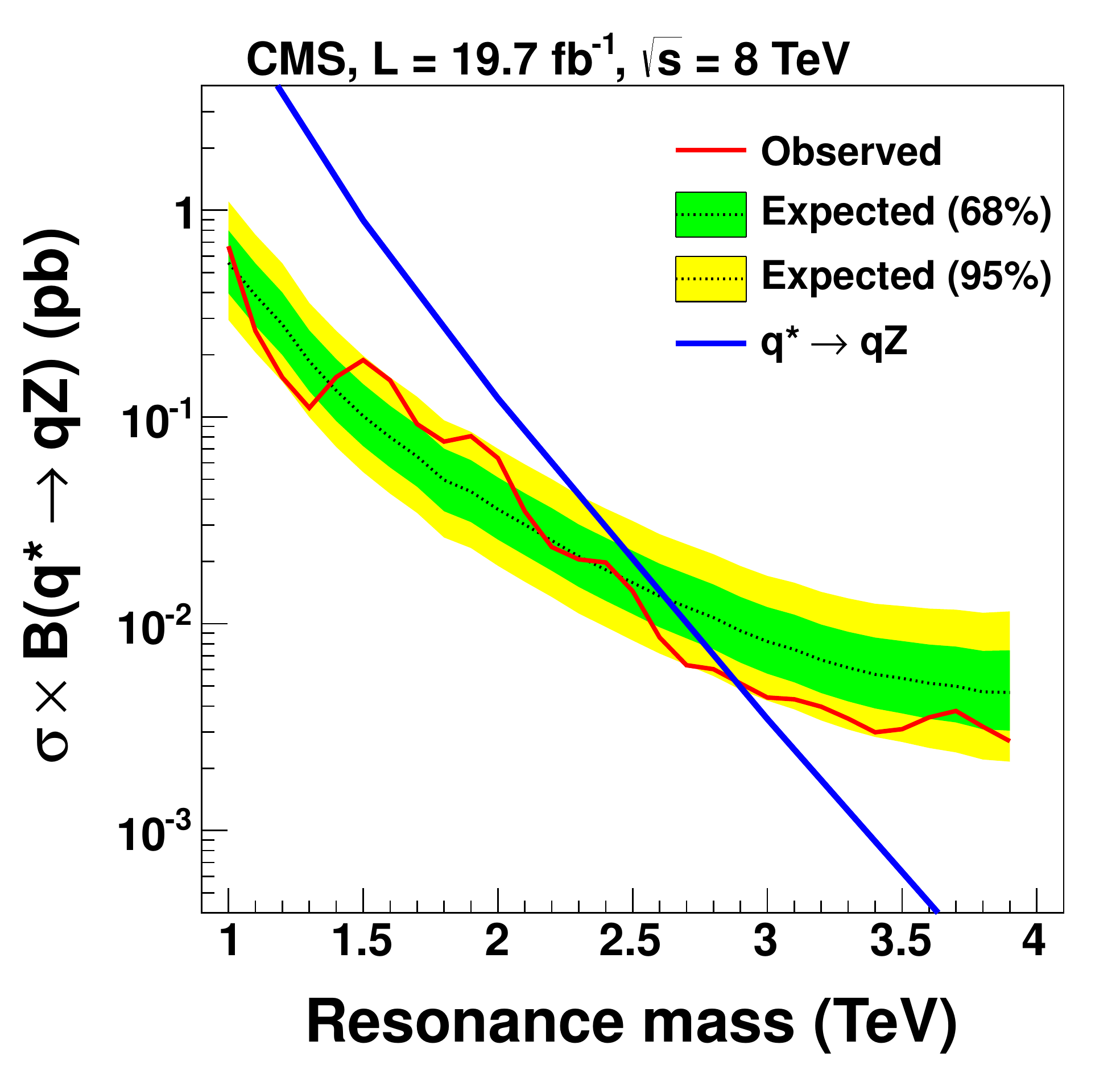}
\includegraphics[width=0.49\textwidth]{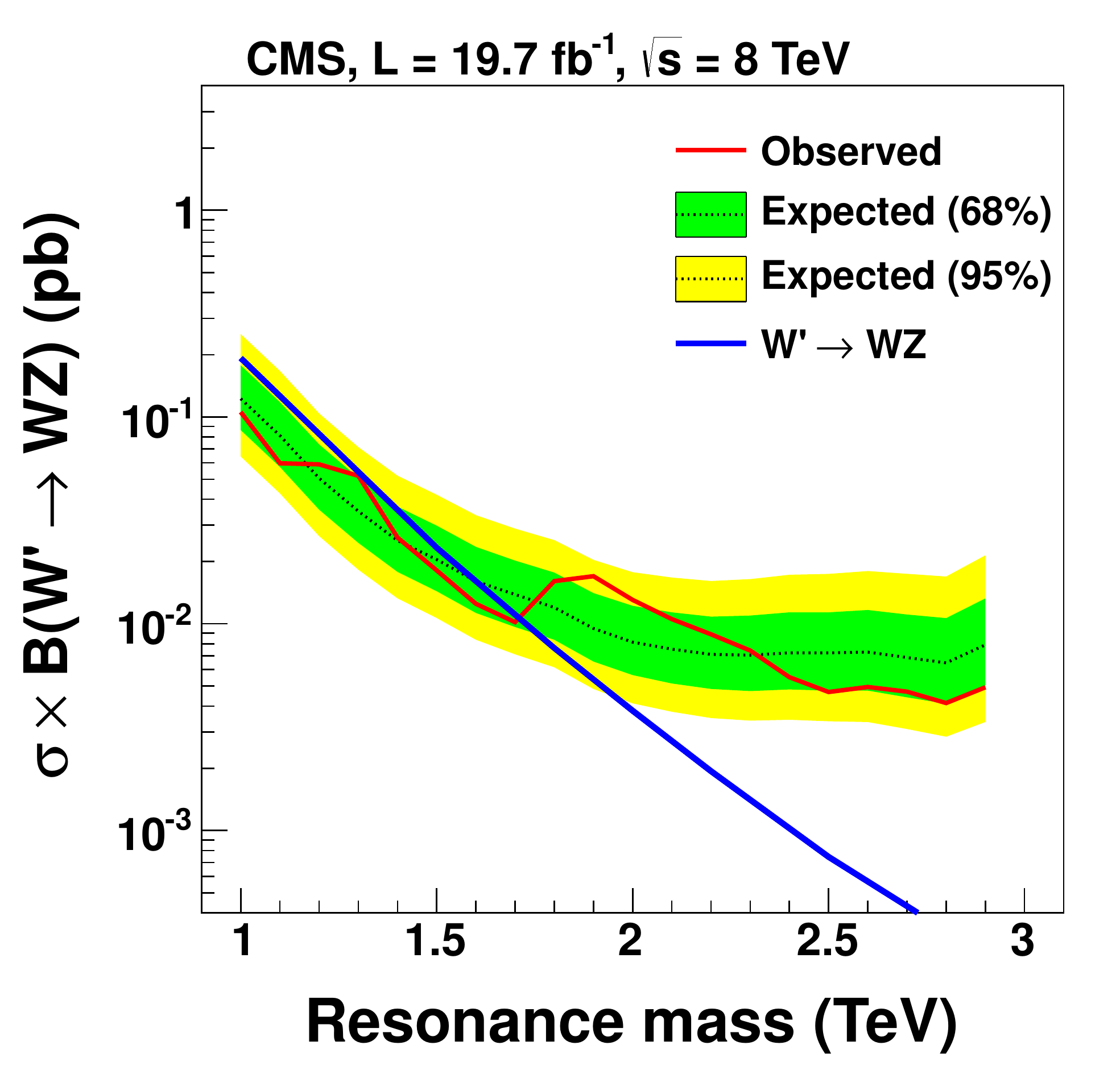}

\caption{Expected and observed 95\% CL limits on the production
  cross section as a function of the resonance mass for (upper left)
  qW resonances, (upper right) qZ resonances, and (bottom) WZ
  resonances, compared to their predicted cross sections for the
  corresponding benchmark models.}
\label{fig:Vtagresults}
\end{figure*}

\begin{figure*}[h!tb]
\centering
\includegraphics[width=0.49\textwidth]{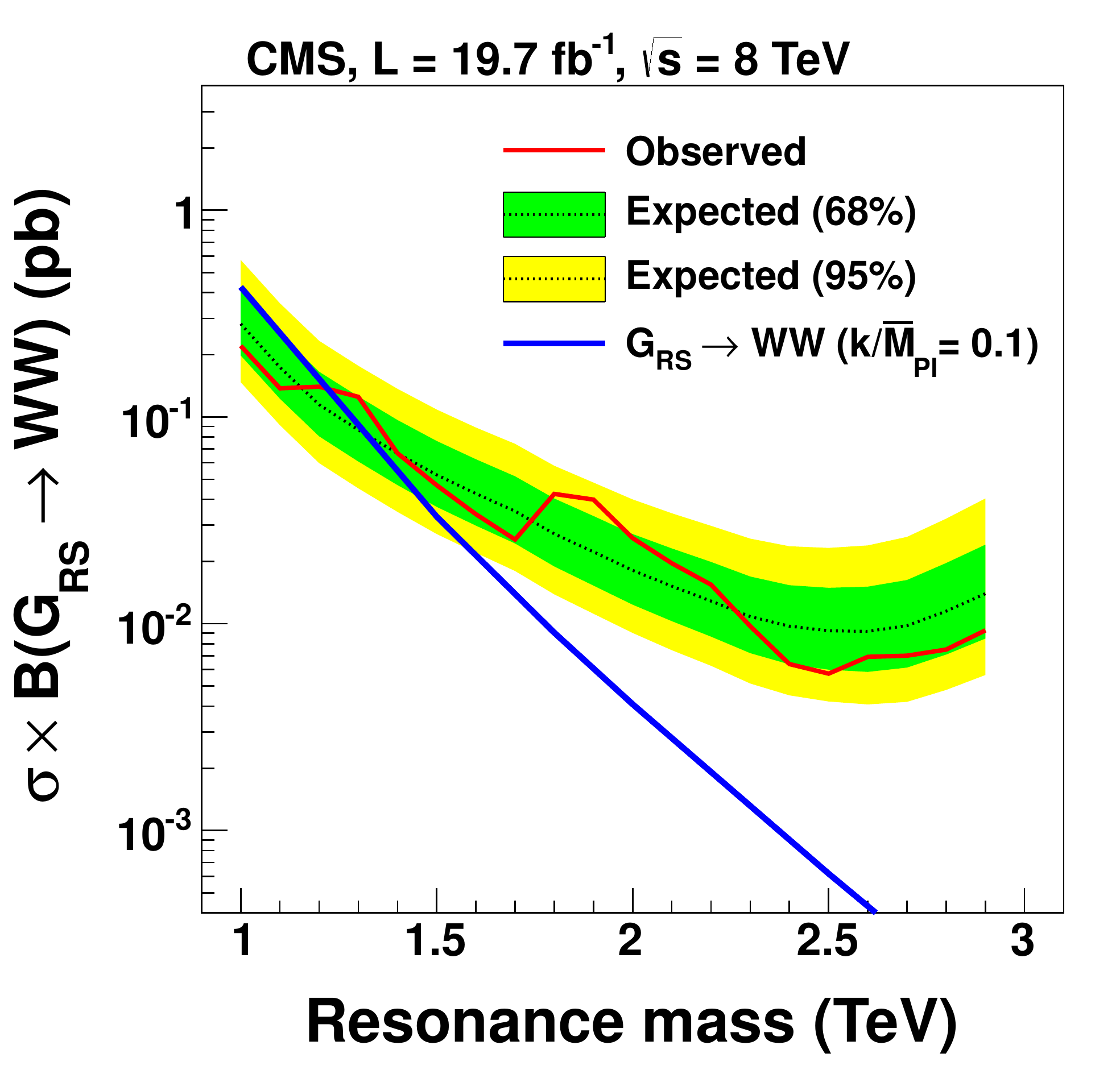}
\includegraphics[width=0.49\textwidth]{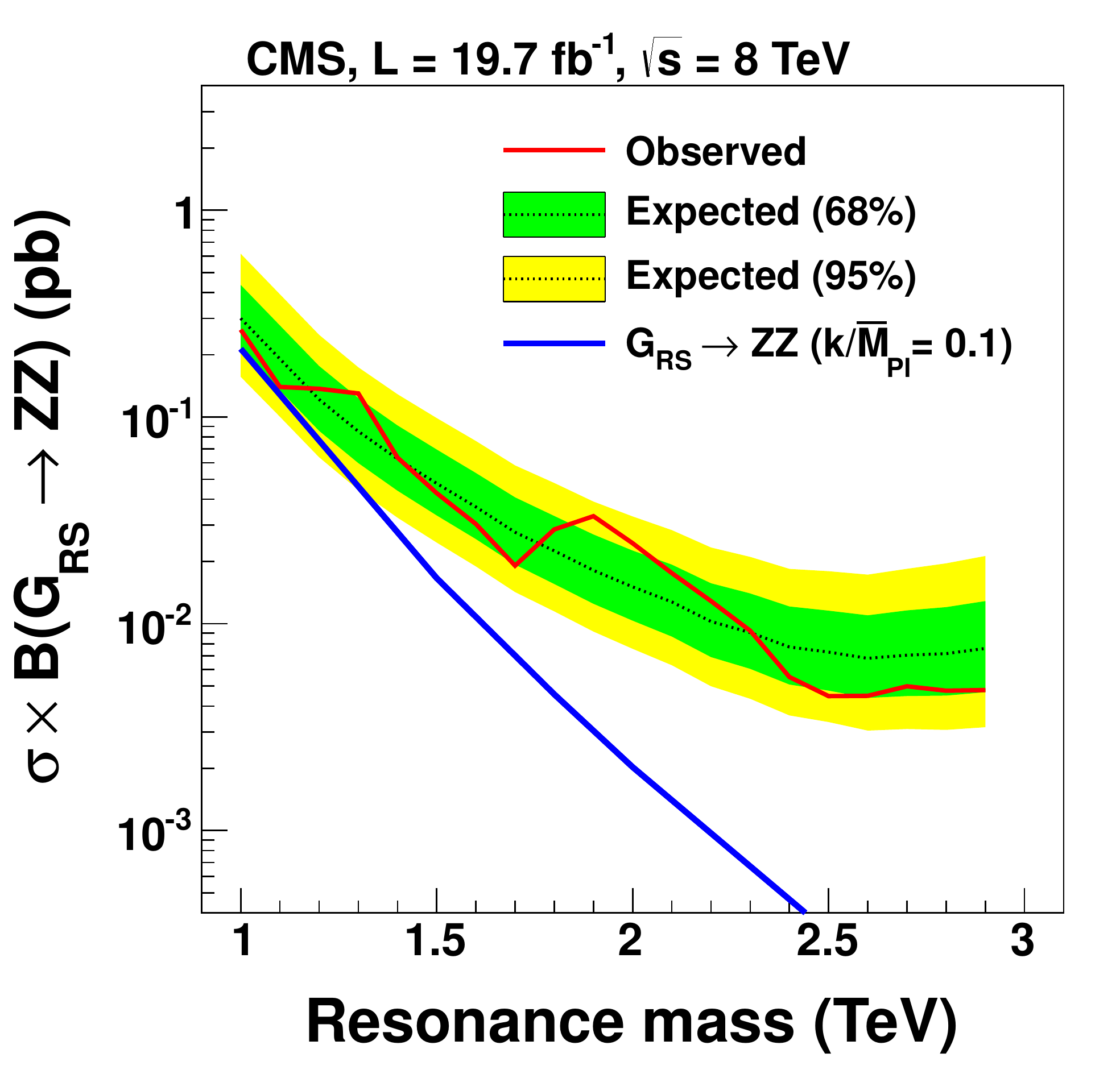}
\includegraphics[width=0.49\textwidth]{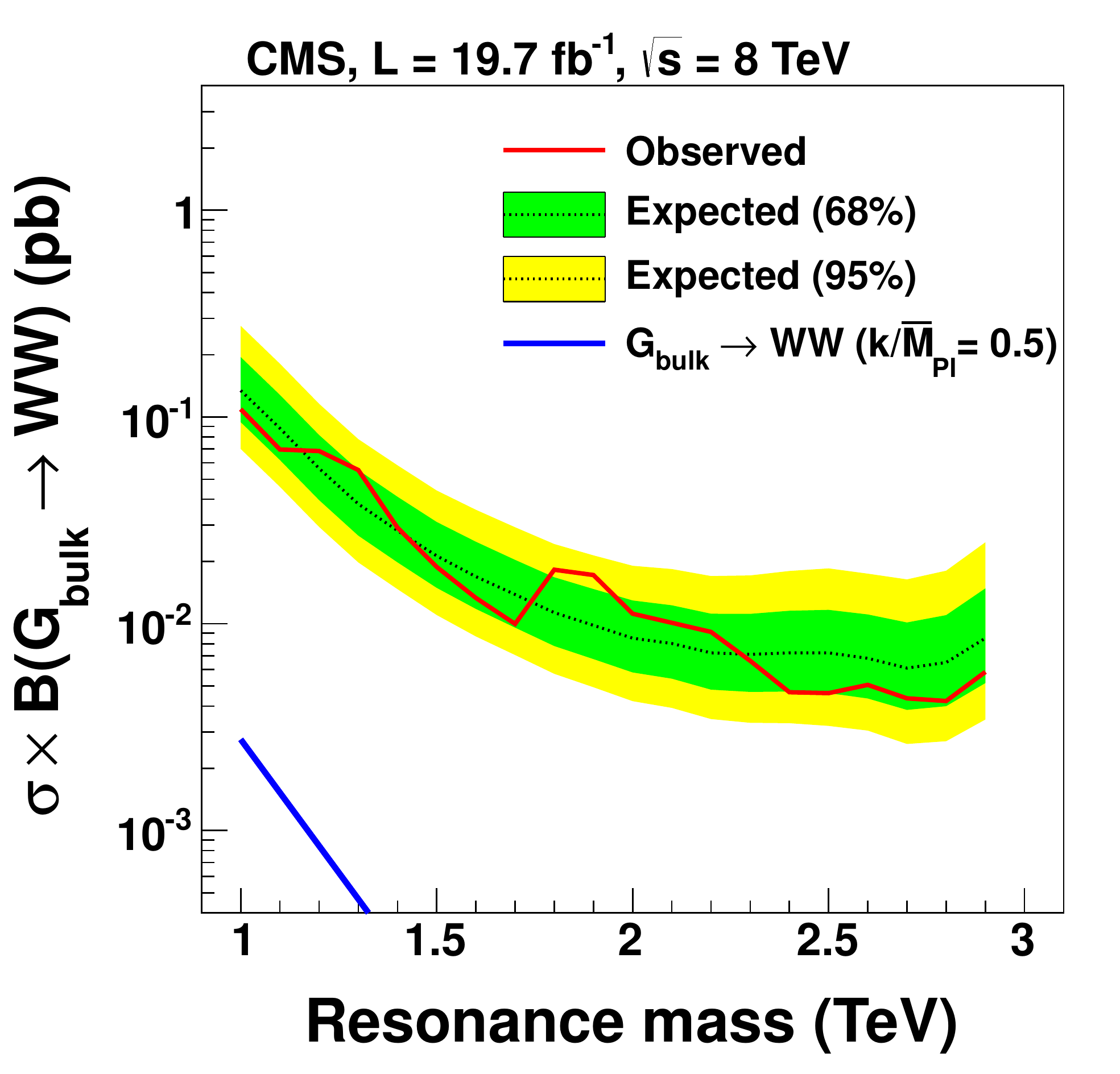}
\includegraphics[width=0.49\textwidth]{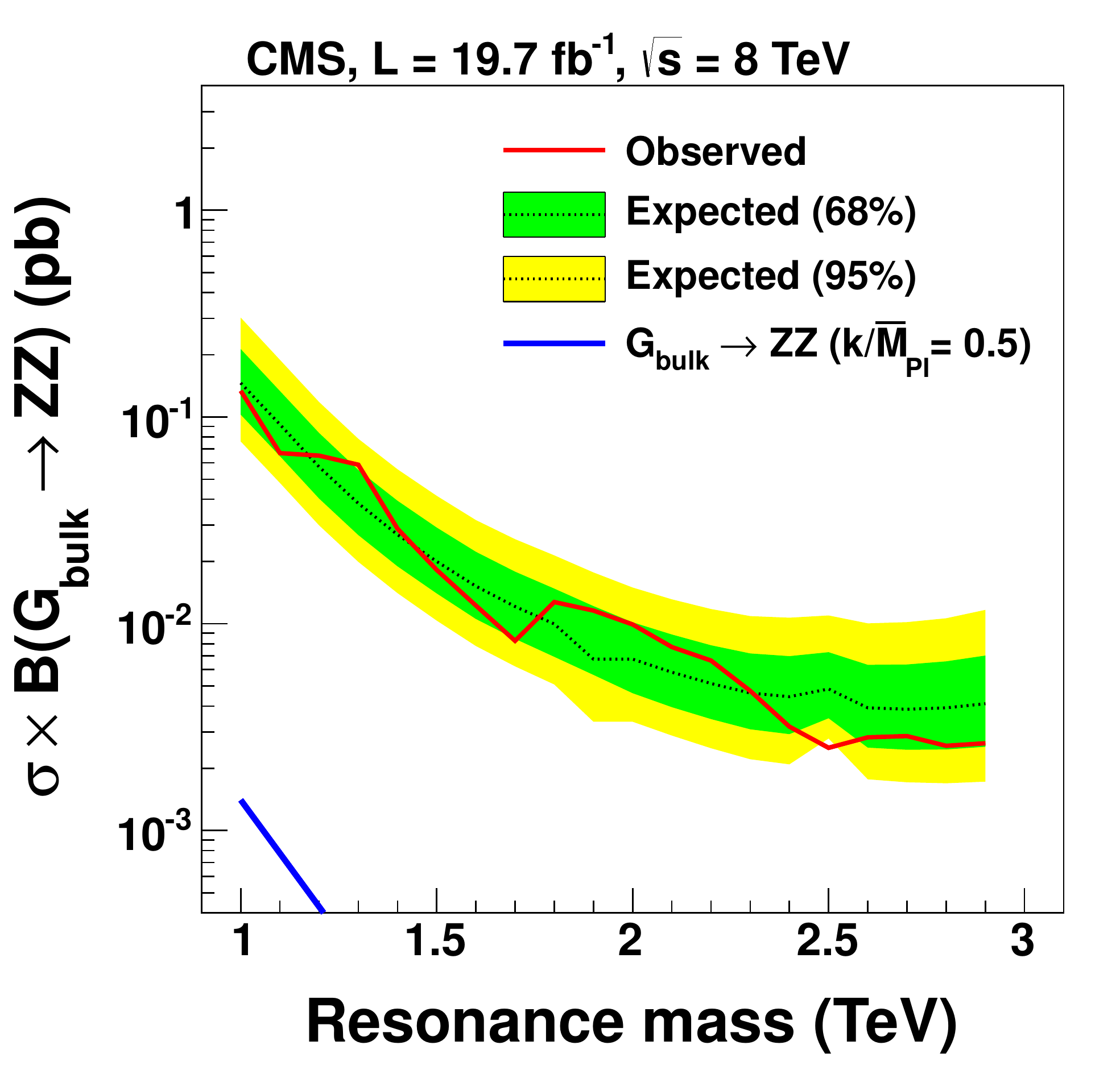}
 \caption{Expected and observed 95\% CL limits on the production cross section
  as a function of the resonance mass for (upper left) $\GRS\to\PW \PW$ resonances,
  (upper right) $\GRS\to\cPZ \cPZ$ resonances, (bottom left) $\GBulk\to\PW \PW$ resonances, and
  (bottom right) $\GBulk \to\cPZ \cPZ$ resonances, compared to the predicted cross sections.
  \label{fig:Vtagresults2}}
\end{figure*}

Figures~\ref{fig:Vtagresults} and~\ref{fig:Vtagresults2} show the
observed and background-only expected
upper limits on the production cross sections for singly and doubly
$\PW/\cPZ$-tagged events, computed at 95\% CL, with the predicted
cross sections for the benchmark models overlaid for
comparison. Table~\ref{table:results} shows the resulting exclusion
ranges on resonant masses. Compared to the previous search in this
channel at $\sqrt{s}=7$\TeVcc~\cite{ref_2011}, the mass limits on
$\Pq^*\to \Pq\PW$ and $\Pq^*\to \Pq\cPZ$ are increased, respectively,
by 0.8 and 0.7\TeV and for the first time mass limits are set on
$\PWpr\to\PW\cPZ$ and $\GRS\to\PW \PW$ models. No mass limits are set
on $\GRS\to\cPZ \cPZ$, $\GBulk\to\PW \PW$ and $\GBulk\to\cPZ \cPZ$,
since the analysis is not sensitive to the small predicted cross
sections.

The systematic uncertainties have minor impact on the limits. The
largest contributions are 5\%, 5\%, and 3\% from $\PW/\cPZ$-tagging
efficiency, JES, and JER, respectively. These numbers are obtained by
quoting the largest change in the observed exclusion limit on the $\GRS
\to\PW\PW$ production cross section, over the entire examined mass
range, when the corresponding uncertainties are removed.

\begin{table}[htb]
\begin{center}
  \topcaption{Summary of observed limits on resonance masses at 95\% CL
    and their expected values, assuming a null
    hypothesis. The analysis is sensitive to resonances heavier than 1\TeV.\label{table:results}}
\begin{tabular}{ ccc}
\hline
Process      & Observed & Expected \\
& excluded mass limit (\TeVns) & excluded mass limit (\TeVns) \\
\hline
$\Pq^*\to \Pq\PW $    & $3.2$  & $3.0$   \\
$\Pq^*\to \Pq\cPZ $    & $2.9$  & $2.6$   \\
$\PWpr\to \PW\cPZ $  & $1.7$  & $1.6$   \\
$\GRS\to \PW \PW $  & $1.2$  & $1.3$   \\
\hline
\end{tabular}
\end{center}
\end{table}

\section{Summary}
\label{sec:conclusions}

An inclusive sample of multijet events corresponding to an integrated
luminosity of \intlumi, collected in pp collisions at
$\sqrt{s}=8$\TeVcc with the CMS detector, is used to measure the
$\PW/\cPZ$-tagged dijet mass spectrum for the two leading jets,
produced within the pseudorapidity range $\abs{\eta} < 2.5$ with a
separation in pseudorapidity of $\abs{\Delta\eta} < 1.3$. The generic
multijet background is suppressed using jet-substructure tagging
techniques that identify vector bosons decaying into \cPaq\Pq' pairs
merged into a single jet. In particular, the invariant mass of pruned
jets and the $N$-subjettiness ratio $\tau_{21}$ of each jet are used
to reduce the initially overwhelming multijet background. The
remaining background is estimated through a fit to smooth analytic
functions. With no evidence for a peak on top of the smoothly falling
background, lower limits are set at the 95\% confidence level on
masses of excited quark resonances decaying into qW and qZ at 3.2 and
2.9\TeV, respectively. Randall--Sundrum gravitons $\GRS$ decaying into
WW are excluded up to 1.2\TeV, and $\PWpr$ bosons decaying into
$\PW\cPZ$, for masses less than 1.7\TeV.  For the first time mass
limits are set on $\PWpr\to \PW\cPZ$ and $\GRS\to \PW\PW$ in the
all-jets final state. The mass limits on $\Pq^*\to \Pq\PW$, $\Pq^*\to
\Pq\cPZ$, $\PWpr\to \PW\cPZ$, $\GRS\to \PW\PW$ are the most stringent
to date.  A model with a ``bulk'' graviton $\GBulk$ that decays into
WW or ZZ bosons is also studied, but no mass limits could be set due
to the small predicted cross sections.

\section*{Acknowledgements}

We congratulate our colleagues in the CERN accelerator departments for
the excellent performance of the LHC and thank the technical and
administrative staffs at CERN and at other CMS institutes for their
contributions to the success of the CMS effort. In addition, we
gratefully acknowledge the computing centres and personnel of the
Worldwide LHC Computing Grid for delivering so effectively the
computing infrastructure essential to our analyses. Finally, we
acknowledge the enduring support for the construction and operation of
the LHC and the CMS detector provided by the following funding
agencies: BMWFW and FWF (Austria); FNRS and FWO (Belgium); CNPq,
CAPES, FAPERJ, and FAPESP (Brazil); MES (Bulgaria); CERN; CAS, MoST,
and NSFC (China); COLCIENCIAS (Colombia); MSES and CSF (Croatia); RPF
(Cyprus); MoER, ERC IUT and ERDF (Estonia); Academy of Finland, MEC,
and HIP (Finland); CEA and CNRS/IN2P3 (France); BMBF, DFG, and HGF
(Germany); GSRT (Greece); OTKA and NIH (Hungary); DAE and DST (India);
IPM (Iran); SFI (Ireland); INFN (Italy); NRF and WCU (Republic of
Korea); LAS (Lithuania); MOE and UM (Malaysia); CINVESTAV, CONACYT,
SEP, and UASLP-FAI (Mexico); MBIE (New Zealand); PAEC (Pakistan); MSHE
and NSC (Poland); FCT (Portugal); JINR (Dubna); MON, RosAtom, RAS and
RFBR (Russia); MESTD (Serbia); SEIDI and CPAN (Spain); Swiss Funding
Agencies (Switzerland); MST (Taipei); ThEPCenter, IPST, STAR and NSTDA
(Thailand); TUBITAK and TAEK (Turkey); NASU and SFFR (Ukraine); STFC
(United Kingdom); DOE and NSF (USA).

Individuals have received support from the Marie-Curie programme and
the European Research Council and EPLANET (European Union); the
Leventis Foundation; the A. P. Sloan Foundation; the Alexander von
Humboldt Foundation; the Belgian Federal Science Policy Office; the
Fonds pour la Formation \`a la Recherche dans l'Industrie et dans
l'Agriculture (FRIA-Belgium); the Agentschap voor Innovatie door
Wetenschap en Technologie (IWT-Belgium); the Ministry of Education,
Youth and Sports (MEYS) of the Czech Republic; the Council of Science
and Industrial Research, India; the Compagnia di San Paolo (Torino);
the HOMING PLUS programme of Foundation for Polish Science, cofinanced
by EU, Regional Development Fund; and the Thalis and Aristeia
programmes cofinanced by EU-ESF and the Greek NSRF.

\bibliography{auto_generated}  % will be created by the tdr script.

\cleardoublepage \appendix\section{The CMS Collaboration \label{app:collab}}\begin{sloppypar}\hyphenpenalty=5000\widowpenalty=500\clubpenalty=5000\textbf{Yerevan Physics Institute,  Yerevan,  Armenia}\\*[0pt]
V.~Khachatryan, A.M.~Sirunyan, A.~Tumasyan
\vskip\cmsinstskip
\textbf{Institut f\"{u}r Hochenergiephysik der OeAW,  Wien,  Austria}\\*[0pt]
W.~Adam, T.~Bergauer, M.~Dragicevic, J.~Er\"{o}, C.~Fabjan\cmsAuthorMark{1}, M.~Friedl, R.~Fr\"{u}hwirth\cmsAuthorMark{1}, V.M.~Ghete, C.~Hartl, N.~H\"{o}rmann, J.~Hrubec, M.~Jeitler\cmsAuthorMark{1}, W.~Kiesenhofer, V.~Kn\"{u}nz, M.~Krammer\cmsAuthorMark{1}, I.~Kr\"{a}tschmer, D.~Liko, I.~Mikulec, D.~Rabady\cmsAuthorMark{2}, B.~Rahbaran, H.~Rohringer, R.~Sch\"{o}fbeck, J.~Strauss, A.~Taurok, W.~Treberer-Treberspurg, W.~Waltenberger, C.-E.~Wulz\cmsAuthorMark{1}
% ...
\vskip\cmsinstskip
% ...
1:~~Also at Vienna University of Technology, Vienna, Austria\\
2:~~Also at CERN, European Organization for Nuclear Research, Geneva, Switzerland\\
% ...

\end{sloppypar}
\end{document}